\documentclass[preprint,12pt]{elsarticle}

\usepackage{lineno,hyperref}
\modulolinenumbers[5]

\usepackage{tikz}
\usepackage{svg}
\usepackage{multirow}
\usepackage{booktabs}
\usepackage{tabularx}
\usepackage{subcaption}
\usepackage{amsmath,amssymb,amsfonts}
\usepackage{algorithmic}
\usepackage{graphicx}
\usepackage{textcomp}
\usepackage{xcolor}
\usepackage{scrextend}

\usepackage{amssymb}
\usepackage{pifont}
\newcommand{\cmark}{\ding{51}}%
\newcommand{\xmark}{\ding{55}}%

\begin{document}

\begin{frontmatter}

\title{A Transfer Learning and Explainable Solution to Detect mpox from Smartphones images}

\author[1]{Mattia Giovanni Campana\corref{cor1}}
\ead{m.campana@iit.cnr.it}

\author[2]{Marco Colussi}
\ead{marco.colussi@unimi.it}

\author[1]{Franca Delmastro}
\ead{f.delmastro@iit.cnr.it}

\author[2]{Sergio Mascetti}
\ead{sergio.mascetti@unimi.it}

\author[2]{Elena Pagani}
\ead{elena.pagani@unimi.it}

\cortext[cor1]{Corresponding author}

\affiliation[1]{
    organization={Institute for Informatics and Telematics of the National Research Council of Italy (IIT-CNR)},
    city={Pisa},
    country={Italy}
}

\affiliation[2]{
    organization={Università degli Studi di Milano, Computer Science Department},
    city={Milan},
    country={Italy}
}

\begin{abstract}
In recent months, the monkeypox (mpox) virus -- previously endemic in a limited area of the world -- has started spreading in multiple countries until being declared a ``public health emergency of international concern'' by the World Health Organization.
The alert was renewed in February 2023 due to a persisting sustained incidence of the virus in several countries and worries about possible new outbreaks.
Low-income countries with inadequate infrastructures for vaccine and testing administration are particularly at risk.

A symptom of mpox infection is the appearance of skin rashes and eruptions, which can drive people to seek medical advice.
A technology that might help perform a preliminary screening based on the aspect of skin lesions is the use of Machine Learning for image classification.
However, to make this technology suitable on a large scale, it should be usable directly on mobile devices of people, with a possible notification to a remote medical expert. 

In this work, we investigate the adoption of Deep Learning to detect mpox from skin lesion images.
The proposal leverages Transfer Learning to cope with the scarce availability of mpox image datasets. As a first step, a homogenous, unpolluted, dataset is produced by manual selection and preprocessing of available image data. It will also be released publicly to researchers in the field. Then, a thorough comparison is conducted amongst several Convolutional Neural Networks, based on a 10-fold stratified cross-validation.
The best models are then optimized through quantization~\cite{Yang_2019_CVPR} for use on mobile devices; measures of classification quality, memory footprint, and processing times validate the feasibility of our proposal.
Additionally, the use of eXplainable AI is investigated as a suitable instrument to both technically and clinically validate classification outcomes.

\end{abstract}



\end{frontmatter}

\section{Introduction}

While the whole world is still dealing with the coronavirus disease (COVID-19) and its mutations~\cite{Callaway_2021}, the recent outbreaks of mpox virus (formerly known as Monkeypox) in different western countries have raised serious concern among public health authorities~\cite{LAI2022787}.
Mpox is a zoonotic disease caused by an orthopoxvirus, and it is closely related with variola (i.e., the smallpox virus), cowpox, and vaccinia viruses~\cite{10.1371/journal.pntd.0010141}.
Although it was first isolated in 1958 from laboratory monkeys, its original hosts also included squirrels, rats, and dormice~\cite{magnus1959pox}.

Since the first human case reported in 1970 in the Democratic Republic of Congo, the spread of mpox was always limited to Central and West Africa, infecting new hosts through close body contact, respiratory droplets, or animal bites, becoming an endemic disease in those regions.
The incubation period ranges from 5 to 21 days, and the actual disease is characterized by generic symptoms such as fever, intense headache and muscle pain, while the most specific sign of mpox is related to the appearance of skin rashes and eruptions that usually begin within 1–3 days of the appearance of fever and tend to be more concentrated on the face and extremities rather than the trunk~\cite{Rizk2022}.

\begin{figure}[t]
    \centering
    \includegraphics[width=\linewidth]{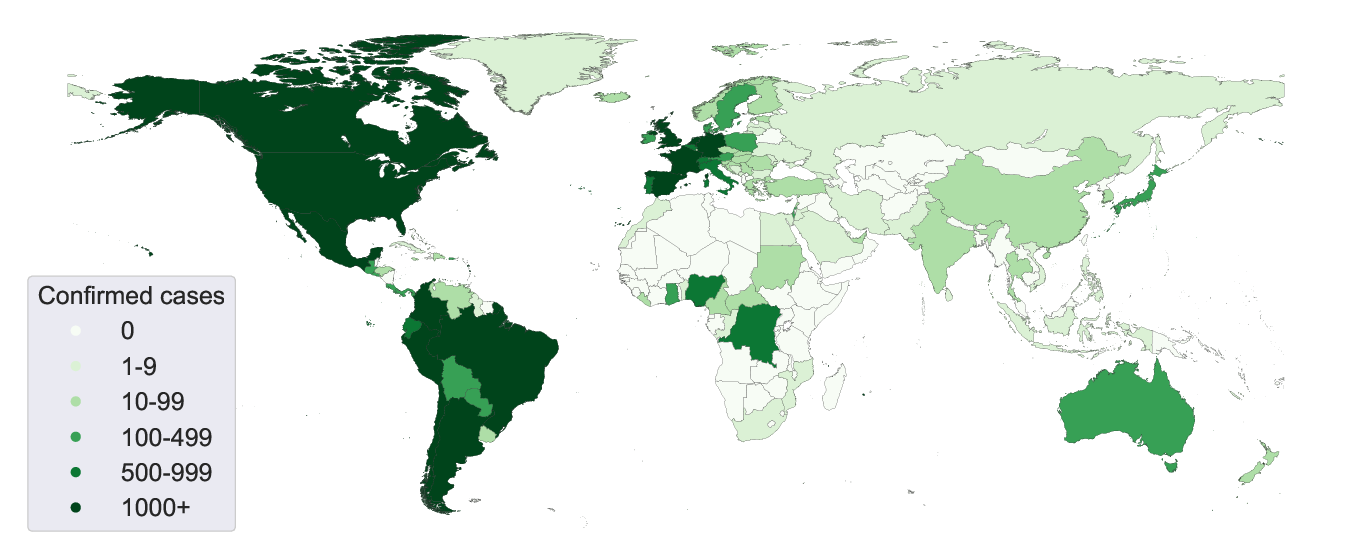}
    \caption{Geographical distribution of the recent mpox outbreak~\cite{WHO-Monkeypox}.}
    \label{fig:mp_geodistribution}
\end{figure}

Since the middle of 2022, a continuously increasing number of cases and sustained chains of transmissions have been reported in regions without direct or immediate epidemiological links to endemic areas, including countries in Europe, North America, and Australia.
On 23 May 2023, the World Health Organisation (WHO) has reported a total of 87,529 laboratory confirmed cases and 1,098 probable cases across 111 countries~\cite{WHO-Monkeypox}, as shown in Figure~\ref{fig:mp_geodistribution}.
Even though mpox is usually not fatal, according to the Centers for Disease Control and Prevention (CDC), people with severely weakened immune systems, children under 1 year old, subjects with a history of eczema, and women who are pregnant or breastfeeding may be more likely to get seriously ill or even die~\cite{CDC-Monkeypox}.

Such rapid and widespread dissemination of the virus has raised several worries in the medical community, highlighting the need for proactive countermeasures in order to prevent another global pandemic~\cite{Rizk2022}.
In this regard, recent studies have emphasized how \emph{mobile-health systems} (\emph{m-health}), along with Artificial Intelligence (AI), can represent a game changer in containing the spread of a virus~\cite{ASADZADEH2021100558, doi:10.1080/10408363.2020.1781779}.
In fact, using the plethora of sensors embedded in modern mobile devices and their increasingly advanced computational capabilities, smartphones, and wearables can be used as low-cost, pervasive, and non-invasive tools to support the early diagnosis of new cases.
For example, Rong et al. developed a smartphone-based fluorescent lateral flow immunoassay for the detection of Zika virus~\cite{RONG2019140}, Brangel et al. proposed the use of a mobile application to read immunochromatographic strips to detect antibodies against Ebola~\cite{Brangel2018}, while more recent works used Deep Learning (DL) models to detect COVID-19 digital biomarkers in respiratory sounds collected by smartphone microphones~\cite{Han2022, 9821076}

In this work, we propose a DL-based m-health solution to detect mpox from skin lesion images captured by personal smartphones.
The considered use case is the following: the user takes a close picture of a skin region that the application uses to automatically detect mpox.
Technically, we use Transfer Learning~\cite{9134370} to adapt state-of-the-art Convolutional Neural Networks (\emph{CNNs}) models~\cite{9451544} to automatically identify visual features of mpox skin rashes, distinguishing the typical symptoms of the virus from skin lesions produced by other pathologies that can be easily confused also by expert eyes, including Chickenpox and Acne, at different severity levels.

Compared with previous works, this paper addresses three issues.
First, the elaboration of available skin lesion images to make them homogeneous with respect to skin section focus and measure, to generate a new homogeneous dataset. In fact, existing datasets include highly heterogeneous images (\textit{e.g.}, images of a group of people or of entire parts of body) that are unsuitable for the considered problem. 

Second, to design a mpox detection system able to run autonomously on personal mobile devices at least to provide a preliminary warning to common users, and that relies on cloud components only for model training and interaction support with a medical expert.  
To this end, we \emph{ optimize} the final DL model to reduce by 4$\times$ the memory footprint of our system, without negatively affecting its classification performance.

Third, to integrate \emph{eXplainable AI} (\emph{XAI}) methods~\cite{10.1007/978-3-030-32236-6_51} to validate the system performance in recognizing the disease from skin lesion pictures and further define a clinical validation process involving medical experts.
According to the literature, XAI techniques greatly improve the general understanding of deep neural networks~\cite{VANDERVELDEN2022102470}, increasing the trust in the overall system by both medical personnel and final users, thus fostering widespread adoption of such digital solutions.
In fact, the target of our proposal is twofold: on the one hand, medical experts can take advantage of such a tool to speed up the diagnosis of new cases, while, on the other hand, final users can autonomously perform a preliminary screening of suspicious skin lesions that must be further investigated by their personal physicians or dermatologists.

In summary, we can highlight our contributions as follows:

\begin{itemize}
    
    \item We adopted Transfer Learning and then fine-tuned 5 state-of-the-art Deep Learning models to detect mpox from skin lesion images.
    
    \item We performed an extensive evaluation of the considered solutions through a series of experiments involving the use of a 10-fold cross-validation technique.

    \item We optimize the best model to be able to perform all the data processing and classification directly on mobile devices, compatibly with the typical memory constraints of commercial smartphones.

    \item We use XAI techniques to validate our model's predictions.

    \item We publicly release all the materials produced in this work, including a curated selection of data called \emph{Mpox Close Skin Images} (\emph{MCSI}) that is composed of $400$ skin images already pre-processed in order to show homogeneous characteristics, which are also perfectly balanced over 4 different classes: mpox, chickenpox, acne, and healthy.

\end{itemize}

The remainder of the paper is organized as follows.
Section~\ref{sec:related} presents the related work regarding the use of Deep Learning in medical images analysis, including preliminary works recently proposed in the literature for the automatic detection of mpox through image processing.
In Section~\ref{sec:proposal}, we describe in detail our mpox detection system for mobile devices.
Section~\ref{sec:experiments} outlines the experimental setup we adopted to evaluate the classification performance of the considered DL models, and discusses the obtained results.
In Sections~\ref{sec:gradcam} and~\ref{sec:optimization} we detail the use of XAI techniques and the mobile-oriented optimization.
Lastly, in Section~\ref{sec:conclusions} we draw our conclusions and present
some directions for future work.

\section{Related Work}
\label{sec:related}
This section first briefly introduces the state of the art in the field of CNN for medical images analysis and in particular in the field of dermatology. Then, it analyzes the existing datasets of mpox skin lesions and the mpox classification techniques. 

\subsection{Convolutional Neural Networks in medical image analysis}
Among the different deep neural networks, Convolutional Neural Networks (CNNs) represent one of the most effective architectures for applications dealing with image data~\cite{8308186}.
CNNs can automatically extract relevant features from raw input images by using a series of convolutional, nonlinear, and pooling layers. Thanks to this characteristic, CNNs made impressive achievements in many computer vision tasks, including image classification, object detection, image segmentation, and face recognition~\cite{9451544}.

In recent years, CNNs have also achieved remarkable results in healthcare applications and, in particular, computer-aided diagnosis.
For example, Majumdara et al.~\cite{MAJUMDAR2023119022} proposed an ensemble of 3 pre-trained CNN models, namely, GoogleNet~\cite{Szegedy_2015_CVPR}, VGG11~\cite{simonyan2014very}, and MobileNetV3Small~\cite{Howard_2019_ICCV} for the detection of breast cancer in histopathological images, obtaining a classification accuracy of $99.16\%$ and $96.95\%$ with two benchmark datasets.
Kumar et al.~\cite{Kumar2020} present a custom CNN model to detect malaria parasites in blood cell images.
Despite the small size of the proposed network (i.e., only $4$ convolutional and pooling layers, followed by $2$ fully connected layers for classification), the authors were able to obtain an accuracy score of $96.62\%$.
Other solutions use CNN as shared feature extractors in multitask models to classify images and, at the same time, localize specific elements of the medical image~\cite{colussi2023ultrasound}.
In the last two years, CNN models have also been adopted in different technological solutions aiming at containing the spread of the COVID-19 pandemic, including systems to monitor social distancing and the use of face masks in public places~\cite{Ansari2021, Singh2021}; to automatically analyze blood samples~\cite{TUNCER2021104579}, chest X-Ray and Computerized Tomography (CT) images~\cite{JIA2021104425}; and to fast screening the population by analyzing respiratory sounds collected from mobile devices, represented as spectrogram images~\cite{Han2022, CAMPANA2023101754}.

Dermatology is another field of application where the use of CNNs is increasingly being investigated~\cite{kassem2021machine}.
Among others, Shetty et al. 
compared the performance of several shallow classifiers (e.g., Random Forest and Support Vector Machines) with a custom CNN in classifying dermoscopic images of different types of skin lesions for skin cancer detection~\cite{Shetty2022}.
The paper shows that, using a dataset of $700$ images and data augmentation techniques, CNN overcame the best classifiers by $7\%$, scoring an overall accuracy of $95.18\%$.
Roy et al.~\cite{8862403} explored several segmentation approaches to detect different skin diseases (e.g., candidiasis and cellulitis).
Finally, Kassem et al.~\cite{9121248} reports $94.92\%$ of accuracy by using transfer learning and a pre-trained GoogleNet to detect melanoma among 8 different classes of skin lesions.


\subsection{Datasets of skin lesions for mpox detection}
\label{sub:dataset}

\begin{table}[t]
\begin{center}
\caption{Mpox datasets}
\label{tab:datasets}
\begin{tabular}{lrrr}
\toprule
\textbf{Name} & \textbf{Images} & \textbf{Classes} & \textbf{Available} \\
\midrule
MSLD~\cite{ali2022monkeypox} & $228$ & 2 & Yes~\footref{msld_link} \\
MDS22~\cite{ahsan2022monkeypox} & $161$ & 4 & No \\
MSID~\cite{diponkor2022monkeypox}  & $770$ & 4 & Yes~\footref{msid_link}  \\
RMSD~\cite{s23041783}  & $2056$ & 2 & Yes~\footref{rmsd_link}  \\
\midrule
MCSI (ours) & $400$ & 4 & Yes \\
\bottomrule
\end{tabular}
\end{center}
\end{table}

Since one of the most common symptoms of mpox is the appearance of skin rashes and lesions, the analysis of skin images is a promising solution for the early detection of this novel global outbreak of the virus. Thus, an annotated dataset of images is required to train the model.

The existing datasets include images of skin lesions caused by mpox as well as images in other classes, for example, images of the skin without lesions or with lesions caused by other diseases.
The four datasets proposed in the literature are summarized in Table~\ref{tab:datasets}.

Ali et al.~\cite{ali2022monkeypox} presented the 
\textit{Monkeypox Skin Lesion Dataset} (\textit{MSLD})\footnote{\label{msld_link}\url{https://www.kaggle.com/datasets/nafin59/monkeypox-skin-lesion-dataset}} that contains $228$ skin lesion images collected from different sources on the Internet and that are divided into two classes: mpox cases and a generic \emph{Others} class, which includes skin lesions caused by other diseases (e.g., Chickenpox and Measles), but also samples without evident lesions.
Ahsan et al.~\cite{ahsan2022monkeypox} provided the  \emph{Monkeypox-dataset-2022} (\textit{MDS22}), which includes a total of $161$ images of mpox, chickenpox, measles and skin without any lesions labeled as \textit{Healthy}.
However, at the time of writing, \textit{MDS22} is no longer accessible.
The third dataset, called \emph{Monkeypox Skin Images Dataset} (\emph{MSID})~\cite{diponkor2022monkeypox} includes $770$ images divided in the same 4 classes as \textit{MDS22}\footnote{\label{msid_link}\url{https://www.kaggle.com/datasets/dipuiucse/monkeypoxskinimagedataset}}.
Finally, the fourth dataset, called \emph{Roboflow Monkeypox Skin Dataset} (\emph{RMSD})~\cite{s23041783} includes $2056$ images divided into 2 classes as \textit{MSLD}. The positive class includes augmented images of mpox, while for the negative case, different images were web scraped for different pathologies such as Lyme, Drug Rash, Pityriasis Rosea Rash, and Ring Worm\footnote{\label{rmsd_link}\url{https://app.roboflow.com/ds/uHWnw424Sk?key=w8YJKfcD2i}}.

Unfortunately, existing datasets have severe limitations.
First, they contain very heterogeneous pictures in terms of both resolution and subjects, including histopathology images, whole-body parts, and full-body images.
For example, one image in MSID represents a group of three people (\texttt{normal184}), others represent people watching in the mirror (\texttt{normal198}) or using skin care product (\texttt{normal188}), while others represent full body parts (\texttt{chickenpox15} or \texttt{monkeypox46}). Similarly, RMSD contains  web-scraped images, that in some cases represent monkeys (\texttt{images109}, \texttt{images224}, \texttt{Monkeypox\_1}), a collage of both normal and positive samples (\texttt{images247}), or images of hospital buildings (\texttt{images17}, \texttt{images49}). In other cases, images are present in the test folder multiple times with different processing or extracted from different sources (\texttt{image9}, \texttt{1\_MONKEYPOX-BOY}, \texttt{image54}).
Also, Altun et al.~\cite{s23041783} take into account skin rushes that are clearly different from those caused by mpox, like those caused by Lyme disease.

While these datasets could possibly be useful for the training of a model aimed at automatically classifying web-scraped images, we argue that they are unsuitable for our use case in which the user takes a close picture of the skin and manually crops it (if needed) so that it only contains the skin rash and not other visual features.
Image heterogeneity may jeopardize the ability of ML models to perform proper classification, due to the presence of ``distracting'' irrelevant features~\cite{AsilianBidgoli2022, 10.1001/jama.2019.18058}.

Another limitation is that images in some classes are under-represented. For example, MSLD contains only $17$ images in the measles class.
This is due to the fact that, for some diseases, there are few images that are public and suitable (\textit{e.g.}, with a sufficient resolution).
However, dealing with such severe imbalanced data poses a challenge for machine learning models, which typically leads to poor predictive performance (especially for the minority class) due to the lack of an equal distribution of training data samples over the different classes.
Although such a problem is usually addressed by oversampling minority class examples or by undersampling the majority class~\cite{10.1007/978-3-030-29407-6_17}, such an approach is unfeasible for small datasets as those publicly available for mpox.
Furthermore, the existing datasets do not include a class of bacterial skin infections (\textit{e.g.}, acne) that, according to the WHO, should be considered in the clinical differential diagnosis of mpox~\cite{WHO-Monkeypox-fact}.


Unlike existing data sets, in this paper we present the \textit{MCSI} (Mpox Close Skin Images) dataset, which includes $400$ homogeneous skin images equally distributed in four classes (\texttt{Mpox}, \texttt{Chickenpox}, \texttt{Acne}, and \texttt{Healthy}).
\textit{MCSI} has been collected by merging other public datasets, and it only includes close skin pictures, as those produced by users taking photos of their own skin lesions from a short distance. 
The details related to both data collection and elaboration are described in Section~\ref{sec:mcsi}.


\subsection{Classification techniques supporting mpox detection}

Three papers propose binary classification techniques trained and evaluated on \textit{MSLD}.
In particular, 
Ali et al.~\cite{ali2022monkeypox}  compare the performance of 3 popular CNN architectures, namely, \textit{VGG16}~\cite{simonyan2014very}, \textit{ResNet50}~\cite{He_2016_CVPR}, and \textit{InceptionV3}~\cite{Szegedy_2016_CVPR}, along with an ensemble of the three with majority voting.
The authors note that \textit{ResNet50} is able to score the best accuracy, $82.96\%$, while the ensemble solution shows a lower performance than the single models.
Sahin et al.~\cite{sahin2022human} use transfer learning and fine-tuning for different CNNs, with the aim of finding the best model to implement on a mobile device.
The experiments show that \textit{MobileNetvV2}~\cite{8578572} obtains the best accuracy ($91.11\%$).
Finally, Alcal{\'a}-Rmz et al.~\cite{10.1007/978-3-031-21333-5_9} present an alternative solution based on GoogleNet that yields $97.08\%$ accuracy.

Other three papers present techniques trained on \textit{MDS22} or \textit{MSID} that hence have the opportunity to distinguish among the four classes defined in these datasets.
Sitaula and Shahi~\cite{sitaula2022monkeypox} present a classifier based on an ensemble of Xception and DenseNet-169 with majority voting. In this case, the technique is trained and evaluated on \textit{MDS22} and it
achieves an accuracy of $87.13\%$.
The paper by Ahsan et al.~\cite{ahsan2022image} also proposes a technique trained and evaluated on \textit{MDS22}.
Specifically, it proposes two studies using a pre-trained version of VGG16: the former classifies mpox versus chickenpox, obtaining $83\%$ accuracy on 18 test images; while the latter obtains $78\%$ accuracy when comparing mpox with all other cases.
Abdelhamid et al.~\cite{abdelhamid2022classification} evaluates optimization algorithms to find the optimal hyperparameters of a deep neural network for image classification in the \textit{MSID} dataset.
The solution reports that GoogleNet yields an accuracy of $98.80\%$.

Finally, ~\cite{s23041783} proposes a classification technique trained on RMSD.
It compares different state-of-the-art models adopting transfer learning in the task of binary classification, obtaining the best results with MobileNetV3 small, with an accuracy of $96.8\%$ and an F-1 score of $0.978$.

Unfortunately, the presented works suffer from a main limitation due to the characteristics of the considered datasets. In addition, they also present some issues from the methodological point of view.
For instance, \cite{sahin2022human} and~\cite{s23041783}) are evaluated with a fixed random split into train-validation-test sets and do not adopt cross-validation.
This approach can lead to overfitting the model on specific dataset partitions, potentially overestimating its generalization capabilities and performance when deployed in real application environments, especially when the training is based on a limited dataset~\cite{vabalas2019machine}.
In other cases, the papers do not clearly specify the evaluation methodology (e.g., ~\cite{abdelhamid2022classification}).
Finally, none of the previous papers present solutions optimized for mobile devices.

In this paper, we train the models with a dataset specifically designed for the addressed problem
such as \textit{MCSI}.
We provide a thorough and reproducible comparison of several state-of-the-art CNNs, 
and we validate the obtained results through the use of Grad-CAM~\cite{Selvaraju_2017_ICCV}, a popular eXplainable-AI technique.

\section{Mpox detection system for mobile devices}
\label{sec:proposal}

\input{fig_proposal2}

Figure~\ref{fig:proposed_system} shows the high-level architecture of the proposed framework to detect mpox from skin lesions images collected from mobile devices.
The whole process can be summarized in two main stages.
In the first stage, we fine-tune a set of pre-trained CNNs on MCSI, and we compare their performance to find the best model for mpox detection, which is afterward optimized for mobile devices.
Fine-tuning includes complex and time-consuming operations and, therefore, is executed on a remote server.

The second stage involves the use of the optimized best-performing model to identify new mpox cases, performing the whole data processing on user devices: a new picture is firstly acquired from the device camera and then cropped in order to contain the target skin lesion. The resulting image is then used as input to the deep learning model that generate the classification.
Moreover, a XAI module is used to both explain and, to some extent, validate the model's prediction, highlighting the most important sections of the input image that led the model output.

In the following, we describe in detail the main building blocks of the proposed solution.


\subsection{Model selection and fine-tuning}
\label{sec:models}
The framework relies on transfer learning to adapt a set of pre-trained CNN to our application scenario, thus reducing the dependence on a large number of training data to build up the target learners~\cite{9134370}.


We consider the following 5 CNNs that represent the state-of-the-art on image classification:

\begin{itemize}

\item \textbf{VGG-16}~\cite{simonyan2014very}, composed by 5 consecutive blocks of convolutional layers for features extraction, followed by 3 fully-connected layers for classification.
Convolutional layers use $3\times3$ kernels with a stride of 1 and padding of 1 to ensure that each activation map retains the same spatial dimensions as the previous layer.
A Rectified Linear Unit (ReLU) activation is performed right after each convolution, and a max pooling operation is used at the end of each block to reduce the spatial dimension. Max pooling layers use $2\times2$ kernels with a stride of 2 and no padding to ensure that each spatial dimension of the activation map from the previous layer is halved.
Finally, two fully-connected layers with $4096$ ReLU activated units are used before a final $1000$ fully-connected softmax layer.

\item \textbf{Inception-Resnet-V2}~\cite{szegedy2017inception} represents a combination of two popular architectures: GoogleNet~\cite{Szegedy_2015_CVPR} and ResNet~\cite{He_2016_CVPR}.
While the former is based on the concept of ``Network in Network''~\cite{lin2013network}, where a large number of convolutional kernels constitute a very deep architecture to increase the network's generalization, the latter introduced the idea of directly bypassing the input information to the output, thus changing the direct learning target value into learning the residual value between the input and the output.
Inception-Resnet-v2 combines the two concepts, using residual
connections instead of filter concatenation, to both accelerate the training and improve the performance.

\item \textbf{NASNetMobile}~\cite{zoph2018learning}, a simplified version of Neural Architecture Search Network (NASNet) proposed by GoogleBrain, which is a scalable CNN architecture consisting of basic building blocks, called \emph{cells}, that are optimized using reinforcement learning.
A cell consists of only a few operations, including both convolutions and pooling, which are repeated multiple times
according to the required capacity of the network.
The mobile
version consists of $12$ cells, with a total of $5.3$ million
parameters.

\item \textbf{MobileNetV3}~\cite{howard2019searching}, a CNN-based architecture especially tuned to best performing on smartphone CPUs through a hardware-aware Network Architecture Search (NAS), combining a series of building-blocks developed by previous models: the depth-wise separable convolutions as an efficient replacement for
traditional convolution layers from MobileNetV1~\cite{howard2017mobilenets}, the linear bottleneck and inverted residual structure introduced by MobileNetV2~\cite{Sandler_2018_CVPR}, and the lightweight attention modules used in MnasNet~\cite{Tan_2019_CVPR}.
The model comes in two flavors - which both are tested in this work - that are \textbf{MobileNetV3-Large} and \textbf{MobileNetV3-Small}, which are targeted for high and low resource use cases, respectively.

\end{itemize}



For all the aforementioned architectures, we take into account their instances pre-trained with ImageNet~\cite{deng2009imagenet}, a large-scale dataset of $3.2$ million images and $1000$ different labels, which is commonly used to train CNNs in the image classification domain~\cite{huh2016makes}.

In order to adapt the models to our new application scenario, we employ the Transfer Learning paradigm by freezing the weights of the features-extraction part and replacing the last fully-connected layers with a novel set of classification layers fine-tuned with our MCSI dataset.
We then validate the considered models through the use of a 10-fold cross-validation procedure and \emph{Hyperband}, 
a broadly used hyperparameter selection algorithm for deep neural networks, which is able to speed up the random search over the parameter spaces through adaptive resource allocation and early-stopping~\cite{li2017hyperband}.
In other words, Hyperband uses a combination of small random searches aimed at partitioning the original search space into smaller sub-spaces.
Once a search iteration is completed, the most promising sub-spaces (i.e., those that allowed the network to obtain the best results) are further explored until a performance plateau is reached or the iterations budget (i.e., the maximum number of iterations) has been exhausted.

\begin{figure}[t]
    \centering
    \includegraphics[width=.9\linewidth]{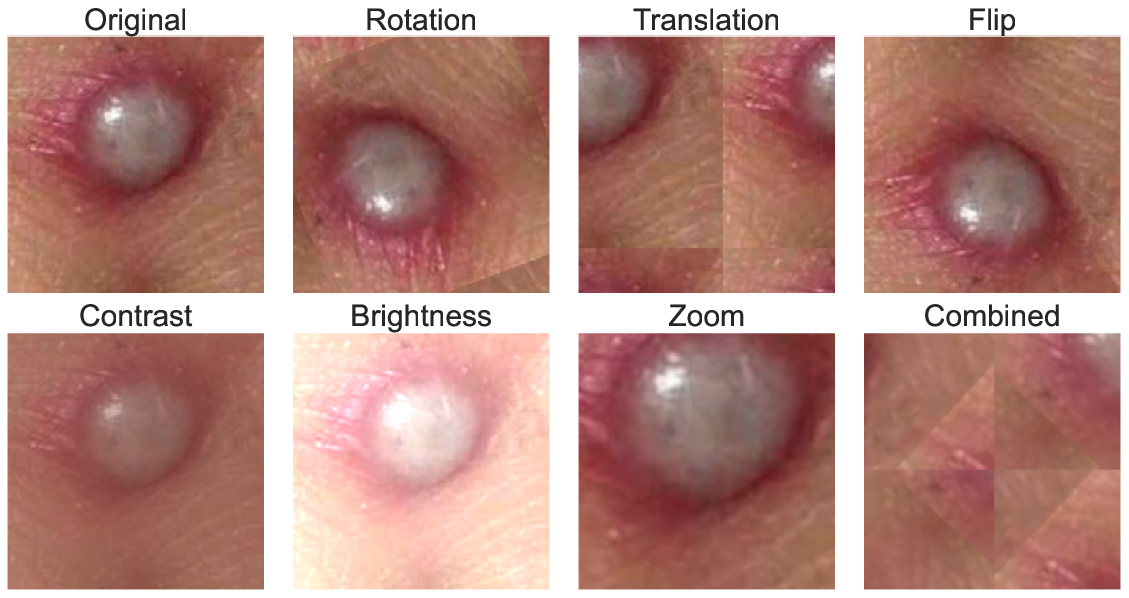}
    \caption{Example of data augmentations used in our experiments.}
    \label{fig:augmentation}
\end{figure}

Moreover, during the evaluation process, we evaluate the feasibility of using data augmentation in our application scenario to possibly improve the performance of the fine-tuned models.
Specifically, we employ the $6$ standard image augmentation techniques~\cite{shorten2019survey} shown in Figure~\ref{fig:augmentation}: (i) \emph{Rotation}, which changes the image angle, simulating different orientations; (ii) \emph{Translation}, simulating different positions of the skin rash inside a specific picture; (iii) \emph{Flip}, which mirrors the image, thus simulating different type of pimples; (iv-v)\emph{Contrast} and \emph{Brightness}, simulating different settings in the amount and intensity of light; and, finally, (v) \emph{Zoom}, scaling the image to simulate variations in the distance between the skin lesion and the smartphone camera.

Data augmentation is not applied to the test and validation sets so as to avoid introducing bias in the models' evaluation.
We include the parameters that affect the augmentation factors (e.g., rotation angle or zoom level) into the tuning phase to identify the set of values that lead to the best classification performance for our application scenario.



\subsection{CNN optimization for mobile devices}

Our main goal is the definition of a mpox detection system that can be entirely executed on mobile devices.
However, neural networks are both computationally and memory intensive.
While modern smartphones are equipped with increasingly powerful hardware (e.g., multicore CPUs and, in some cases, dedicated GPUs) that allows performing the inference phase in just a few milliseconds, neural models' size still represents a challenge, making it difficult to deploy them on embedded systems with limited memory resources.

To cope with this issue, several techniques have been recently proposed to reduce the memory footprint of deep learning models, including \emph{pruning}, where redundant connections among hidden units are removed, or \emph{weight clustering}, which consists in replacing similar weights in a layer with a representative value found by clustering algorithms~\cite{NIPS2015_ae0eb3ee, han2015deep}.
\emph{Quantization} is another practical and broadly used technique to optimize deep learning models by simply lowering the operations' precision from $32$-bit floats to $16$-bit floats or even $8$-bit integers.
Despite its simplicity, it is generally effective in reducing the overall model's size by $4\times$ at least, with little or no degradation in terms of accuracy~\cite{8927153}.
Furthermore, while other approaches must be used during the training phase, quantization can be applied to the final fine-tuned model yield by transfer learning.


\subsection{Explaining the model's predictions}

Deep learning models including CNNs are weak in explaining their inference process and final predictions, thus being typically considered as a black-box.
This characteristic is not suitable for many real-world applications, and especially for the health sector, in which explainability and transparency are essential not just for researchers and developers to validate their models, but also for the users who can be directly affected by AI decisions.

For this reason, increasing attention has recently been paid to eXplainable AI (XAI) techniques with the aim of making AI models more transparent, understandable, and interpretable, so as to increase trust in their predictions.
Different XAI approaches have been recently proposed for deep learning models, based on the characteristics of specific architectures~\cite{10.1007/978-3-030-32236-6_51}.
According to Ibrahim et al.~\cite{10.1145/3563691}, XAI techniques for CNNs can be categorized as \emph{decision models} and \emph{architecture models}.
While the former solutions aim at identifying the parts of an image that mostly contributed to the network decision, the latter explores the network internals, analyzing the mechanism of both hidden layers and neurons.

Given its simplicity in both implementation and interpretability, for our mpox detection system, we decided to use Grad-CAM~\cite{selvaraju2017grad} as XAI approach, one of the most popular decision models used in medical imaging~\cite{singh2020explainable, bourdon2021explainable}.
Grad-CAM is defined as an importance attribution feature algorithm that generates a visual explanation for class-discriminative prediction.
More specifically, it captures the features that positively influence the prediction of a given class, by computing its gradient and then propagating it back to the last convolutional layer to finally generate a heatmap that visually represents the most relevant part of the input image that has led the model to that prediction.
As a preliminary stage, this approach represents a useful tool to validate the ability of the considered fine-tuned deep models in correctly detecting mpox.
Then, after a thorough clinical validation performed by experts with a larger amount of data, such a XAI technique can be also implemented on the mobile device of the final user to support the pre-screening of suspicious skin lesions.





\section{Experimental evaluation}
\label{sec:experiments}

In this section, we present the experimental evaluation performed to identify the best DL model.
We first describe the \textit{MCSI} dataset.
Then, we describe in detail the evaluation protocol and metrics adopted to measure the classification performances of the fine-tuned CNN models.
Finally, we discuss the obtained results. 
The source code and data are publicly available on our Github repository\footnote{\url{https://github.com/mattiacampana/Monkeypox-Detection}} while the cross-validation folds are completely reproducible by the provided code.

\subsection{The Mpox Close Skin Images dataset}
\label{sec:mcsi}
The Mpox Close Skin Images (\textit{MCSI}) dataset has been created according to three design principles. First, the dataset only includes close skin images with or without skin lesions, as these are representative of the pictures that can be collected by the users in the considered use case.
Second, \textit{MCSI} contains images of skin lesions caused by diseases that, according to the WHO, should be considered in the mpox clinical differential diagnosis~\cite{WHO-Monkeypox-fact}.
In particular, we consider one class for chickenpox rash and one for acne, which is a common skin condition caused by bacterial skin infections.
Third, the number of samples should be balanced among the different classes, to avoid a bias.

\begin{figure}[t]
    \centering
    \includegraphics[width=.9\linewidth]{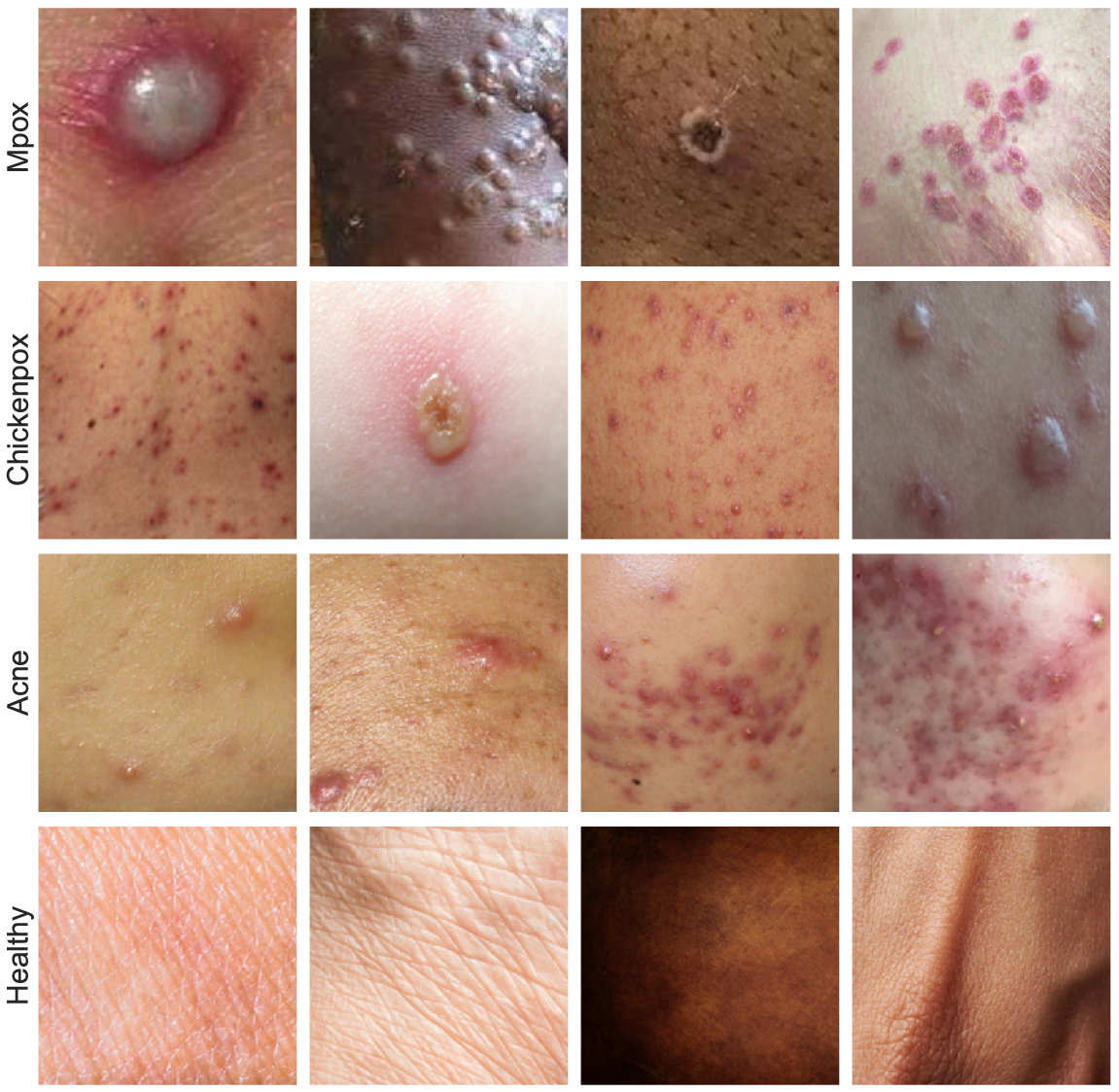}
    \caption{Sample images from the collected dataset for each of the 4 considered classes: \emph{Mpox}, \emph{Chickenpox}, \emph{Acne}, and \emph{Healthy}.}
    \label{fig:dataset_samples}
\end{figure}

Specifically, \textit{MCSI} includes: (1) images of \textbf{Mpox} cases collected by Ali et al.~\cite{ali2022monkeypox} by web scraping news portals, publicly available case reports, and websites; (2) pictures of \textbf{Chickenpox} lesions available on the Hardin Library for the Health Sciences of the University of Iowa\footnote{\url{http://hardinmd.lib.uiowa.edu/chickenpox.html}}, (3) samples of \textbf{Acne} at different severity levels, collected by Wu et al.~\cite{Wu_2019_ICCV} and freely available on Github\footnote{\url{https://github.com/xpwu95/LDL}}, and (4) samples of skin without evident lesions, named as \textbf{Healthy}, available in the dataset collected by Mu{\~n}oz-Saavedra et al.~\cite{munoz4186534monkeypox}.

Based on the aforementioned data sources, we manually selected $100$ images for each of the $4$ considered categories in order to create a balanced dataset of skin pictures that best match our application scenario.
Specifically, we discarded full-body images and pictures with marks drawn by the medical personnel to clearly indicate the skin lesion on the patient's body, and we cropped the pictures to focus on the specific skin lesion.
Figure~\ref{fig:dataset_samples} shows a few samples of the images contained in our dataset for each label.

\subsection{Evaluation protocol and metrics}
\label{sec:eval_protocol}

The evaluation protocol is based on the following:
we decided to rely on the \emph{10-fold stratified cross-validation} approach to avoid biasing the results based on specific train/validation/test splits of the dataset.
The procedure can be summarized as follows.
Firstly, we partition the dataset into 10 folds, ensuring that all the considered classes of images are equally represented in each fold.
For each of the 10 cross-validation iterations, one fold is selected as the \emph{test set}, while the remaining 9 represent the \emph{development set} that is further divided into stratified non-overlapping \emph{train} ($75\%$) and \emph{validation} ($25\%$).
We apply data augmentation at run-time, only on the training sets.
Then, a hyperparameters tuning process (Section~\ref{sec:hyper}) is used by training models on the train set and testing them on the validation set.
The model yielding the best performance is then tested on the test set, providing the performance for that iteration.

We measure the average performance of the fine-tuned models obtained during the 10-fold cross-validation by using the different base models as backbone for features extraction, and a set of fully-connected layers are trained from scratch for classification.
We consider the following standard classification metrics: \emph{Accuracy}, which is the percentage of correct predictions; 
\emph{Sensitivity}, which represents the true positive rate; 
\emph{Specificity}, that indicates the true negative rate; 
and \emph{F-1 Score}, which is the harmonic mean of Precision and Sensitivity.

We perform the whole process for two different classification settings: binary and multiclass.
In the former, we evaluate the models' ability to identify mpox cases without distinguishing the other classes, which are merged into a single ``other'' class.
In the latter, the models learn to distinguish all the four classes available in MCSI.

Furthermore, we conduct a statistical analysis to determine the level of significance in the obtained classification results in terms of accuracy, thereby identifying the most effective model(s) for our specific application scenario.
Initially, we examine the outcomes of the two classification tasks without employing data augmentation.
We conduct this analysis by using \emph{Repeated Measures Analysis of Variance (ANOVA-RM)}, a statistical method used to assess significant differences among the means of three or more dependent groups.
We chose this method because our models were evaluated on the same data folds, making the results dependent on each other.
Moreover, even though ANOVA is generally robust to slight deviations from normality assumptions (especially with small sample sizes), we use the \emph{Shapiro-Wilk} test to assess the distribution characteristics of the results. This evaluation aimed to confirm that the models' results can be approximated by a normal distribution.
Since ANOVA-RM only indicates the presence or absence of a significant difference, without specifying the specific groups that differ from each other,
we subsequently employ the \emph{Tukey's Honest Significant Difference (HSD)} test, which allows us to determine the significance of performance differences between each pair of models, providing a more detailed understanding of the disparities.

Next, we perform a statistical assessment to evaluate the impact of data augmentation on each model, by employing the following procedure.
The initial step involves using the Shapiro-Wilk test to determine whether the performance of the model, both with and without augmentation, follows a normal distribution.
If both distributions pass the test (i.e., $p > 0.05$), we proceed to assess their homoscedasticity using \emph{Bartlett's} test, which determines if the distributions have equal variances. However, if either distribution failed the Shapiro-Wilk test, indicating non-normality, we utilize the non-parametric \emph{Wilcoxon's rank-sum} test as an alternative to the two-sample t-test.
Finally, if the distributions exhibited homoscedasticity, we employ the standard \emph{Independent t-test} to evaluate their statistical significance; otherwise, we use the \emph{Corrected Independent t-test} (also known as \emph{Welch's test}) instead.

\subsection{Hyperparameters tuning}
\label{sec:hyper}

\begin{figure}[t]
    \centering
    \includegraphics[width=.92\textwidth]{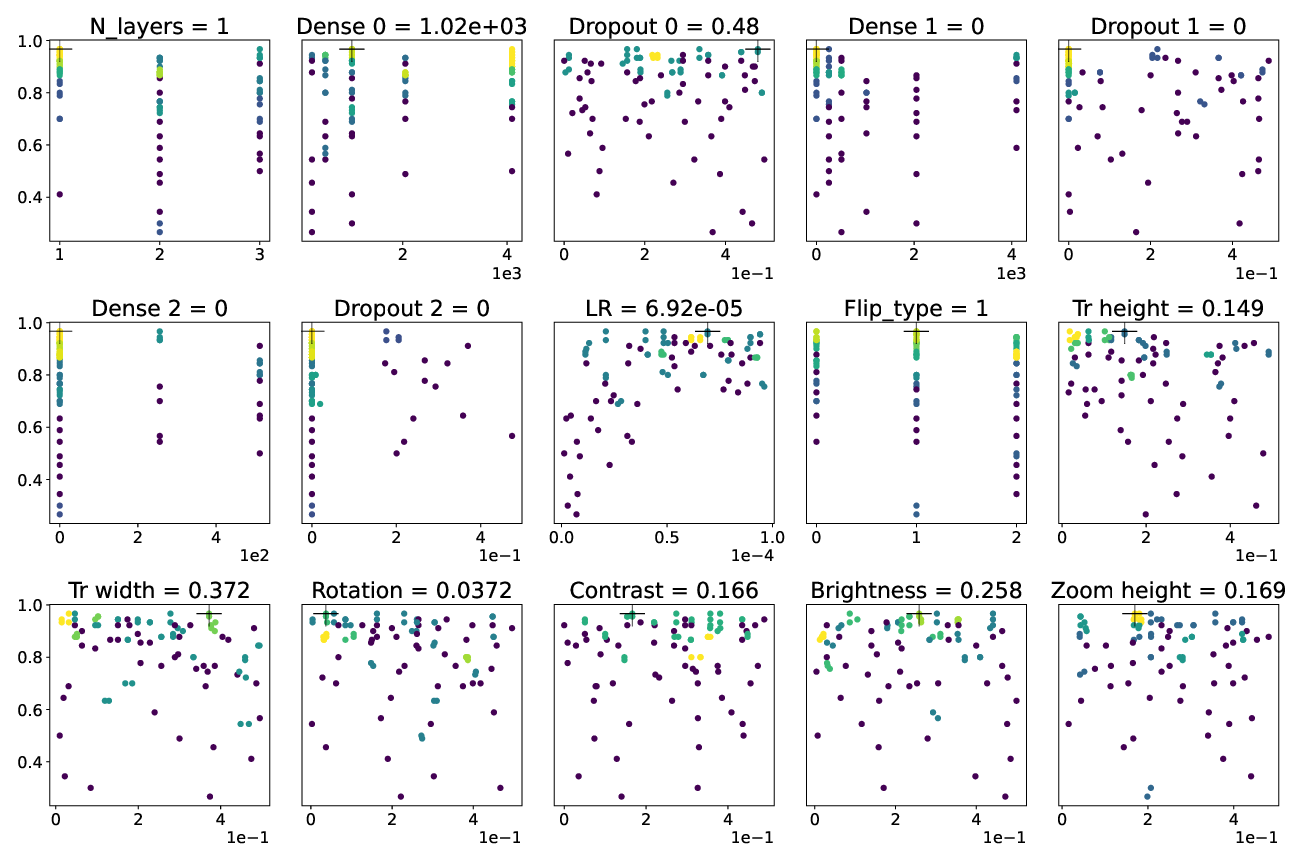}
    \caption{Explored parameters for MobileNetV3Large with augmentation (on fold 0)}
    \label{fig:exploration}
\end{figure}

Actual performances of deep neural networks depend on several hyperparameters that must be tuned in order to find the best configuration for every application scenarios.
We adopted Hyperband for fine-tuning the model and data augmentation parameters.
Considering the model's parameters, we tune
the \emph{learning rate} (\texttt{LR} in the range $\left[1e-6, 0.001\right]$) and the \emph{number of classification layers} (\texttt{N\_layers} among values $\left\{1, 2, 3\right\}$).
Then, for each classification layer, we tune the \emph{number of hidden neurons} (\texttt{Dense} among the values $\left\{256, 512, 1024, 2048, 4096\right\}$) and the \emph{dropout} rate (\texttt{Dropout} in the range $\left[0, 0.5\right]$).


Regarding the data augmentation, we explore two different types of parameters' spaces: continuous and discrete.
The former is defined within $\left[0, 0.5\right]$ and governs the application of \texttt{Rotation}, \texttt{Zoom}, \texttt{Contrast}, \texttt{Brightness}, \texttt{Translation} (both horizontally, \texttt{Tr-width}, and vertically, \texttt{Tr-height}), indicating the percentage in which each operation is applied on the original image.
For example, the value $0.2$ for \texttt{Rotation}, represents a random rotation of the image between $\left[-20\%, +20\%\right]$).
The latter controls the application of \texttt{Flip type}, which may be applied in three different modalities:  \textit{Vertical} ($0$), \textit{Horizontal} ($1$), and the combination of the two ($2$).

Figure \ref{fig:exploration} shows an example of the parameters space explored by Hyperband during the fine-tuning of MobileNetV3Large with data augmentation.
The X-axis indicates the exploration space for a given parameter and can include a finite set of values (\textit{e.g.}, the \texttt{N\_layers}) or can be continuous in a given interval (\textit{e.g.}, \texttt{Dropout}). 
Instead, Y-axis indicates the accuracy levels.
In order to ease the visualization, the density of points is shown with colors (with the \textit{viridis} color map): a single point is shown in purple while multiple overlapping points are shown in yellow.
Finally, the cross symbol ($+$) highlights the combination of parameters that produced the best results, which is also reported on the sub-plot titles.
Note that the parameters \texttt{Dense} and \texttt{Dropout} refer to the corresponding classification layer.
So, for example, \texttt{Dense 1} represents the number of hidden neurons in classification layer 1.
Hence, if a classification layer does not exist (as in the case of layer 2 when \texttt{N\_layers} is 2) the corresponding \texttt{Dense} and \texttt{Dropout} parameters have a value of zero.

\subsection{Mpox detection performances}
\label{sec:results}

In this section, we present in detail the results obtained by fine-tuning the considered CNN architectures
in both binary and multiclass classification settings, with and without data augmentation.
We also present an analysis of their ability to correctly represent image data samples in the latent features space, thus providing additional support to the standard evaluation metrics.

\subsubsection{Binary classification task}

\begin{table}[t]
\caption{Binary classification performance of the considered base models, with and without data augmentation in the training phase. The performance are reported as mean and standard deviation over the 10-folds of the cross-validation.}
\label{tab:binary_classification_results}
\centering
\resizebox{\textwidth}{!}{%
    \begin{tabular}{lcrrrrrr}
    \toprule
    Base model & Augmentation & Accuracy & Sensitivity & Specificity & F-1 Score \\
    \midrule
    \multirow{2}{*}{VGG16}
        & \xmark  & .898 $(\pm.059)$  & .833 $(\pm.106)$ & .710 $(\pm.223)$ & .847 $(\pm.099)$ \\
        & \cmark  & .890 $(\pm.028)$  & .835 $(\pm.027)$ & .730 $(\pm.067)$ & .847 $(\pm.032)$ \\
    \midrule
    \multirow{2}{*}{InceptionResNetV2}
        & \xmark  & .732 $(\pm.051)$  & .568 $(\pm.063)$ & .240 $(\pm.196)$ & .548 $(\pm.093)$ \\
        & \cmark  & .728 $(\pm.068)$  & .544 $(\pm.109)$ & .180 $(\pm.244)$ & .515 $(\pm.132)$ \\
    \midrule
    \multirow{2}{*}{NASNetMobile}
        & \xmark  & .811 $(\pm.038)$  & .726 $(\pm.061)$ & .550 $(\pm.151)$ & .732 $(\pm.055)$ \\
        & \cmark  & .835 $(\pm.044)$  & .727 $(\pm.080)$ & .510 $(\pm.173)$ & .744 $(\pm.081)$ \\
    \midrule
    \multirow{2}{*}{MobileNetV3Small}
        & \xmark  & $\mathbf{.930 (\pm.041)}$  & .877 $(\pm.067)$ & $\mathbf{.780 (\pm.123)}$ & $\mathbf{.898 (\pm.061)}$ \\
        & \cmark  & .921 $(\pm.043)$  & .872 $(\pm.062)$ & $\mathbf{.780 (\pm.114)}$ & .886 $(\pm.060)$ \\
    \midrule
    \multirow{2}{*}{MobileNetV3Large}
        & \xmark  & .930 $(\pm.042)$  & .861 $(\pm.086)$ & .730 $(\pm.177)$ & .889 $(\pm.077)$ \\
        & \cmark  & $\mathbf{.930 (\pm.039)}$ &  $\mathbf{.878 (\pm.071)}$ & $\mathbf{.780 (\pm.140)}$ & .889 $(\pm.060)$ \\
        
    \bottomrule
    \end{tabular}
}
\end{table}

\begin{figure}[t]
    \centering
    \begin{subfigure}[]{0.9\linewidth}
        \centering
        \includegraphics[width=\linewidth]{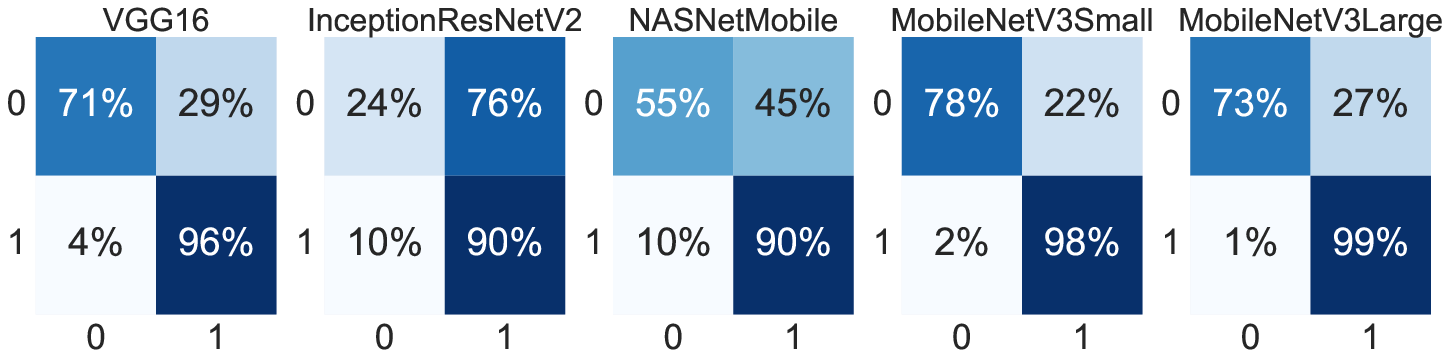}
        \caption{}
        \label{fig:cm_binary_normal}
    \end{subfigure}\\
    \begin{subfigure}[]{0.9\linewidth}
        \centering
        \includegraphics[width=\linewidth]{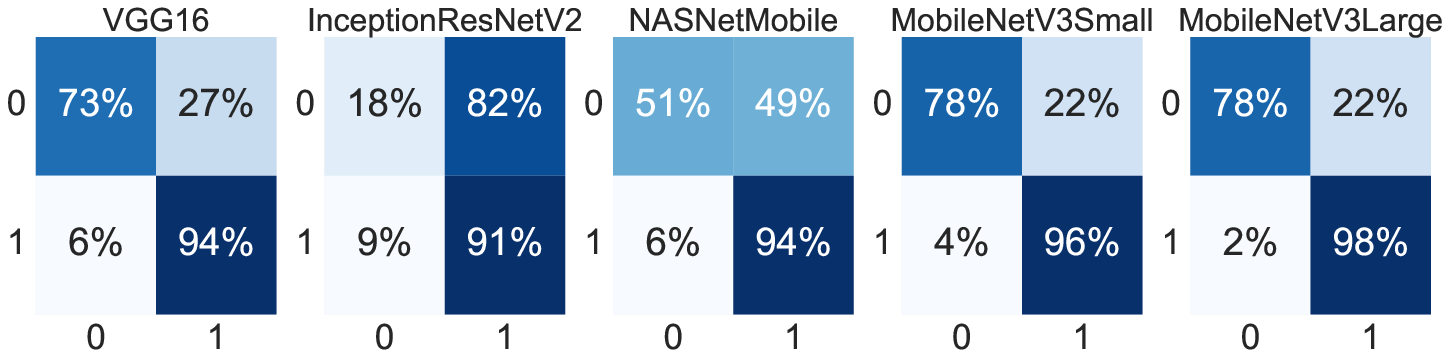}
        \caption{}
        \label{fig:cm_binary_augmented}
    \end{subfigure}\\
    \caption{Confusion matrices related to the binary classification task with original training data (a) and by employing data augmentation (b). Label 0 refers to \texttt{Mpox} samples, while label 1 indicates the generic class \texttt{Others}.}
    \label{fig:cm_binary}
\end{figure}

Table~\ref{tab:binary_classification_results} summarizes the binary classification results of the fine-tuned models, both with and without data augmentation; the results are expressed in terms of mean and standard deviations of the considered evaluation metrics, calculated over the 10-folds of the cross-validation.

Most of the considered base models are able to reach an accuracy level above $80\%$.
Yet, InceptionResNetV2 performs worst,
thus clearly indicating that such an architecture is not able to detect mpox skin rashes from lesions produced by other pathologies.
This is even clearer by observing the confusion matrix in Figure~\ref{fig:cm_binary}, noting that the model incorrectly classifies $76\%$ of the overall mpox samples with the original training data and $82\%$ with data augmentation.

NASNetMobile obtains better results than InceptionResNetV2, but its specificity score is still too low, and its misclassification rate is particularly high to be considered a valid candidate for our system.
On the other hand, VGG16 is capable of far better performance than the previous models.
In this case, we can also note a small improvement introduced by using data augmentation, reducing the percentage of incorrectly classified mpox samples from $29\%$ to $27\%$.

The two variants of MobileNetV3 obtain the best results, reaching in both cases an average accuracy level of $0.93$ and with comparable results for all the considered metrics.
MobileNetV3Small is able to reach the maximum value also in terms of F-1 score, overcoming by approximately $10\%$ the performance of the larger model.
In terms of the rate of misclassification without data augmentation, MobileNetV3Small improves MobileNetV3Large by $5\%$, while the larger model performs slightly better in classifying data samples labeled \texttt{ Others}.
On the other hand, in this case, data augmentation seems to introduce more confusion in the model predictions.
In fact, while it allows MobileNetV3Large to improve its \texttt{Mpox} detection rate, at the same time, it increases the misclassification of \texttt{Others} samples for both models, reaching an error rate of $4\%$ and $2\%$ for MobileNetV3Small and MobileNetV3Large, respectively.

Despite MobileNetV3 achieving the highest classification score, the statistical analysis does not reveal significant differences in accuracy compared to VGG16, with a probability of $p=0.609$.
On the contrary, the analysis confirms that InceptionResNetV2 is the least performing model, exhibiting lower performance compared to the other architectures.
It shows a decrease of $-16.5\%$ compared to VGG16 ($p=0.0$), a decrease of $-8\%$ compared to NASNetMobile ($p=0.004$), and a decrease of $-19.5\%$ compared to the two MobileNetV3 alternatives ($p=0.0$).

Finally, regarding the utilization of data augmentation, the statistical analysis verifies that employing this technique does not significantly impact the average performance of the models, obtaining probabilities considerably higher than the significance threshold of $0.05$ for all the architectures.
Specifically, we observe a probability of $p=0.625$ for VGG16, $p=0.857$ for InceptionResNetV2, $p=0.226$ for NASNetMobile, $p=0.602$ for MobileNetV3Small, and no difference at all for MobileNetV3Large, obtaining a probability of $p=1.0$.

\subsubsection{Multiclass classification task}

\begin{table}[t]
\caption{Multiclass classification performance of the considered base models, with and without data augmentation in the training phase. The performance are reported as mean and standard deviation over the 10-folds of the cross-validation.}
\label{tab:classification_results}
\centering
\resizebox{\textwidth}{!}{%
    \begin{tabular}{lcrrrrrr}
    \toprule
    Base model & Augmentation & Accuracy & Sensitivity & Specificity & F-1 Score \\
    \midrule
    \multirow{2}{*}{VGG16}
        & \xmark  & .779 $(\pm.052)$  & .779 $(\pm.054)$ & .927 $(\pm.018)$ & .777 $(\pm.057)$ \\
        & \cmark  & .745 $(\pm.059)$  & .744 $(\pm.059)$ & .915 $(\pm.020)$ & .738 $(\pm.062)$ \\
    \midrule
    \multirow{2}{*}{InceptionResNetV2}
        & \xmark  & .396 $(\pm.087)$  & .398 $(\pm.088)$ & .780 $(\pm.023)$ & .388 $(\pm.084)$ \\
        & \cmark  & .301 $(\pm.057)$  & .301 $(\pm.067)$ & .767 $(\pm.023)$ & .252 $(\pm.078)$ \\
    \midrule
    \multirow{2}{*}{NASNetMobile}
        & \xmark  & .464 $(\pm.073)$  & .464 $(\pm.073)$ & .822 $(\pm.025)$ & .461 $(\pm.076)$ \\
        & \cmark  & .504 $(\pm.103)$  & .505 $(\pm.104)$ & .835 $(\pm.034)$ & .499 $(\pm.106)$ \\
    \midrule
    \multirow{2}{*}{MobileNetV3Small}
        & \xmark  & .846 $(\pm.062)$  & .847 $(\pm.061)$ & .948 $(\pm.020)$ & .843 $(\pm.065)$ \\
        & \cmark  & .859 $(\pm.054)$  & .860 $(\pm.052)$ & .954 $(\pm.017)$ & .860 $(\pm.049)$ \\
    \midrule
    \multirow{2}{*}{MobileNetV3Large}
        & \xmark  & $\mathbf{.882 (\pm.057)}$ & $\mathbf{.881 (\pm.055)}$ & $\mathbf{.960 (\pm.019)}$ & $\mathbf{.879 (\pm.058)}$ \\
        & \cmark  & .866 $(\pm.088)$  & .866 $(\pm.080)$ & .956 $(\pm.029)$ & .863 $(\pm.086)$ \\
    \bottomrule
    \end{tabular}
}
\end{table}
\begin{figure}[t]
    \centering
    \begin{subfigure}[]{\linewidth}
        \centering
        \includegraphics[width=\linewidth]{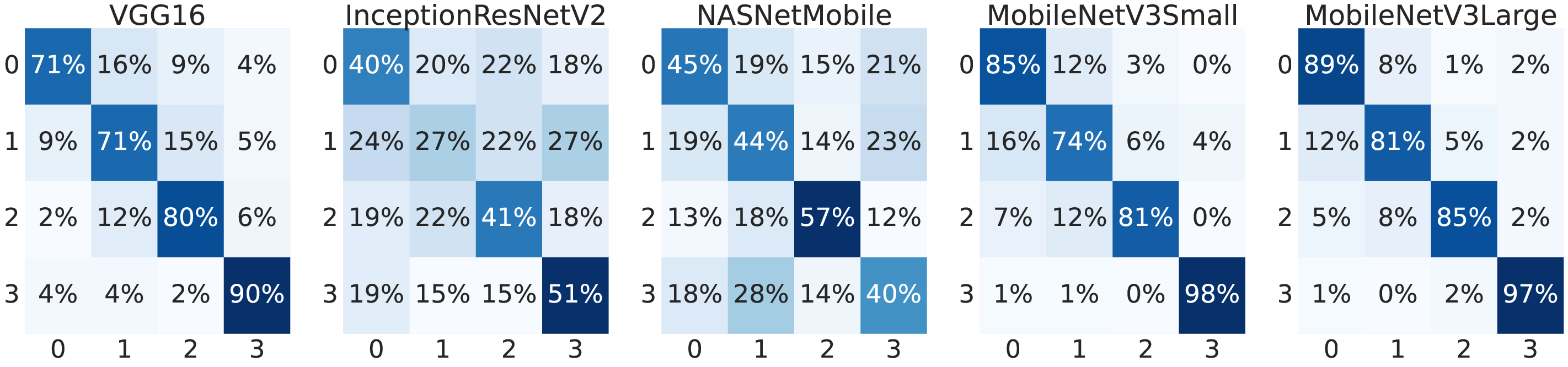}
        \caption{}
        \label{fig:cm_multiclass_normal}
    \end{subfigure}\\
    \begin{subfigure}[]{\linewidth}
        \centering
        \includegraphics[width=\linewidth]{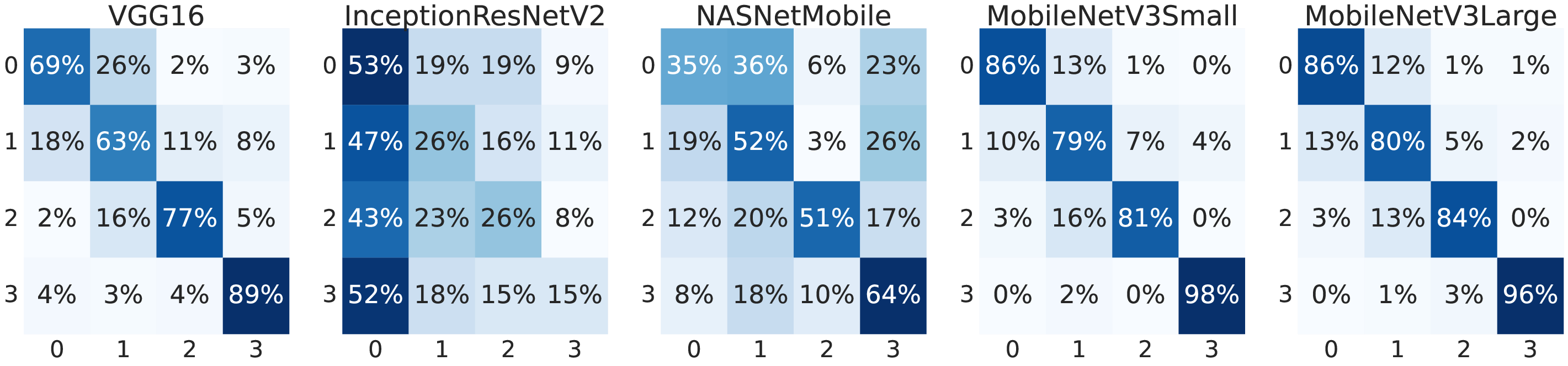}
        \caption{}
        \label{fig:cm_multiclass_augmented}
    \end{subfigure}
    \caption{Confusion matrices related to the multiclass classification setting with original training data (a) and with data augmentation (b). Label 0 refers to Acne samples, label 1 indicates Chickenpox, label 2 indicates \emph{Mpox}, while label 3 indicates the normal class.}
    \label{fig:cm_multiclass}
\end{figure}

Table~\ref{tab:classification_results} summarizes the multiclass classification results of the fine-tuned models, again with and without data augmentation, over the 10-fold cross-validation.
It is worth knowing that the specificity in the multiclass setting is the average of the specificity for each class.
More specifically, for a given class $C$, we calculate the specificity of the model based on the one-vs-all approach, thus as the binary problem of distinguishing between samples belonging to $C$ (positive samples) and samples in all other classes (negative samples).
Specificity is calculated as true negative, the number of negative cases that are correctly identified as negative, divided by true negatives plus false positives, which is the number of negative cases that are incorrectly identified as positive.

Similarly to the binary results, InceptionResNetV2 and NASNetMobile show the worst performances, 
clearly indicating their inability to recognize the different pathologies in the images.
Moreover, data augmentation further reduces the performance of InceptionResnetV2, reducing its F-1 score to $0.252$, while it boosts the F-1 score of NASNetMobile to $0.499$. In Figure~\ref{fig:cm_multiclass} we can note in detail how these two models wrongly classify each class and, in particular, how InceptionResNetV2 tends to classify every sample as \texttt{Acne} (i.e., class $0$).
In contrast, VGG16 yields better results,
although similarly to InceptionResnetV2 data augmentation slightly decreases its performance.

The MobileNetV3 variants achieve the best results also in the multiclass setting.
The MobileNetV3Small yields slightly lower performance: $-3.6\%$ in accuracy, $-3.4\%$ and $-1.2\%$ for sensitivity and specificity and $-3.6\%$ in terms of F-1 score.
On the other hand, it benefits more from data augmentation, improving its F-1 score from $0.843$ to $0.860$.
Quite the opposite happens for MobileNetV3Large; in fact, with data augmentation, all its indexes drop.
Nevertheless, the confusion matrices clearly show how both of the MobileNetV3 variants are able to successfully identify samples in the \texttt{Mpox}, \texttt{Acne}, and \texttt{Healthy} classes (almost $98\%$ of accuracy, both for augmented and non-augmented models), while \texttt{Chickenpox}  represents the hardest class, where MobileNetV3Small scores an accuracy of $79\%$ by augmenting the training data, and the larger variant reaches $80\%$ and $81\%$, respectively with and without data augmentation.

Statistical analysis generally confirms the classification results obtained in our study.
Indeed, there were no significant differences found between InceptionResNetV2 and NASNetMobile ($p=0.188$), which both perform worse than the other considered models.
Furthermore, the two variations of MobileNetV3 exhibited a very high probability of $p=0.776$, suggesting that there were no significant differences between them.

In contrast to the binary classification problem, in the multiclass setting, a noticeable difference can be observed between MobileNetV3Large and VGG16 ($p=0.0154$), while MobileNetV3Small and VGG16 are similar with a probability of $0.2189$.
This difference can be attributed to the fact that in the two-sample tests among the three models, the performance of MobileNetV3Small fell between the other two. Indeed, on average, it showed a slight decrease of $3.6\%$ in accuracy compared to its larger variant, while performing better than VGG16 by $6.5\%$.

Finally, in the case of data augmentation, most of the models did not show statistically significant differences. The probabilities observed were $p=0.190$ for VGG16, $p=0.330$ for NASNetMobile, $p=0.551$ for MobileNetV3Small, and $p=0.734$ for MobileNetV3Large. Only InceptionResNetV2 showed a probability below the threshold at $p=0.012$, confirming the largest drop in performance of $6.5\%$ in terms of accuracy.

To sum up, we can consider both the MobileNetV3 variants as the best choice to detect mpox from skin lesion images, while the larger model is preferable to accurately distinguish mpox from similar diseases.
Moreover, based on the statistical analysis, we can also note that data augmentation does not lead to significant performance improvements, highlighting the need for a larger amount of original training data, as well as a further investigation of more sophisticated approaches of image data augmentation.

\begin{figure}[t]
    \centering
    \begin{subfigure}[]{.3\linewidth}
        \centering
        \includegraphics[width=\linewidth]{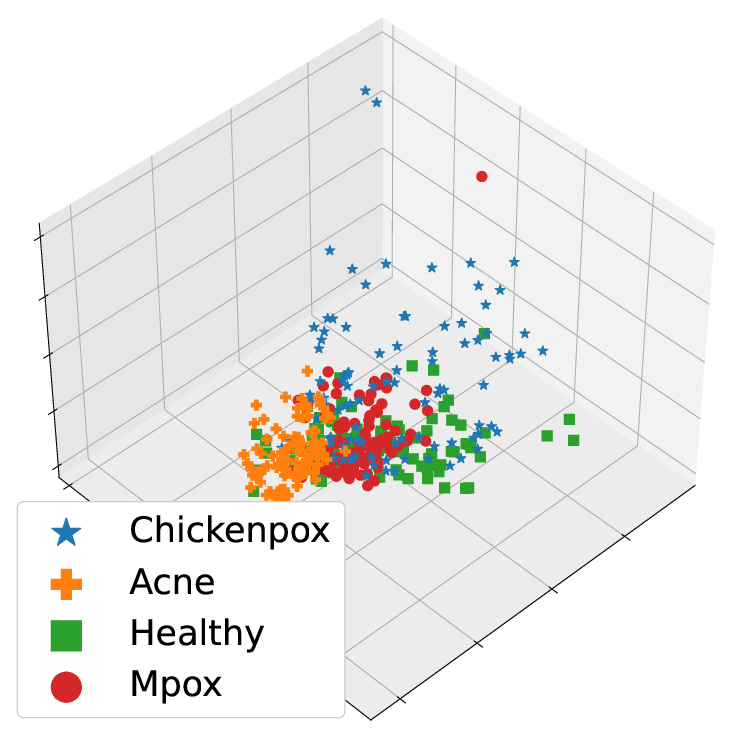}
        \caption{VGG16}
    \end{subfigure}
    \begin{subfigure}[]{.3\linewidth}
        \centering
        \includegraphics[width=\linewidth]{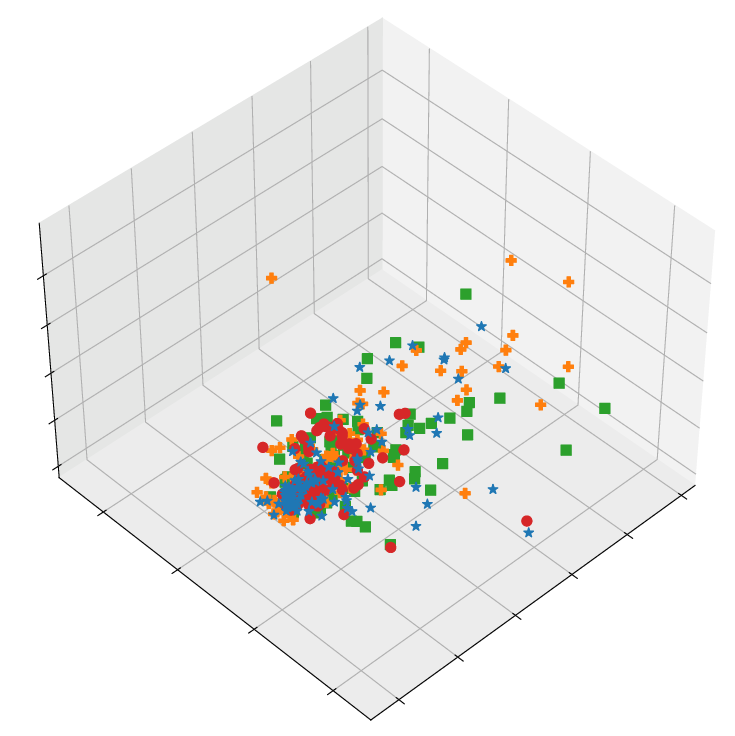}
        \caption{InceptionResNetV2}
    \end{subfigure}
    \begin{subfigure}[]{.3\linewidth}
        \centering
        \includegraphics[width=\linewidth]{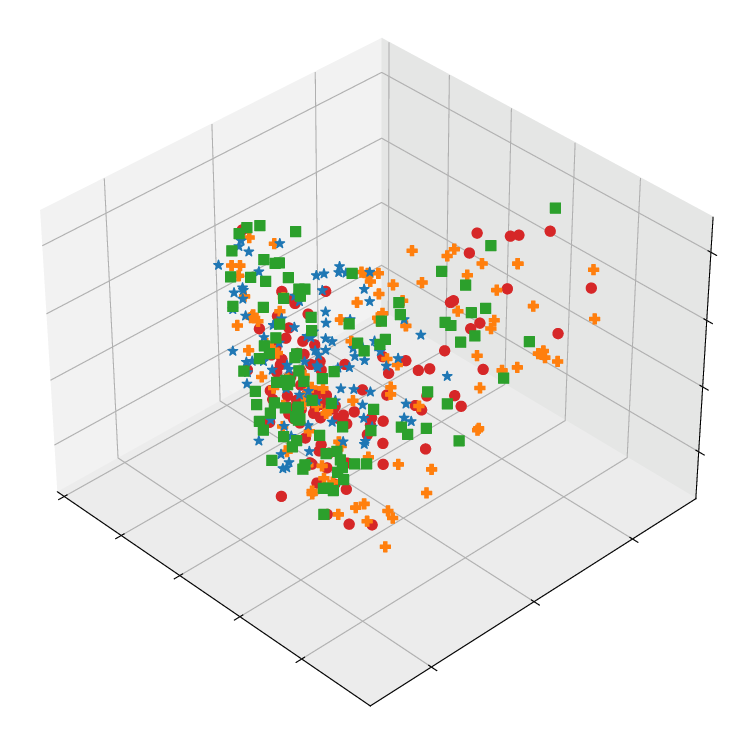}
        \caption{NASNetMobile}
    \end{subfigure}\\
    \begin{subfigure}[]{.3\linewidth}
        \centering
        \includegraphics[width=\linewidth]{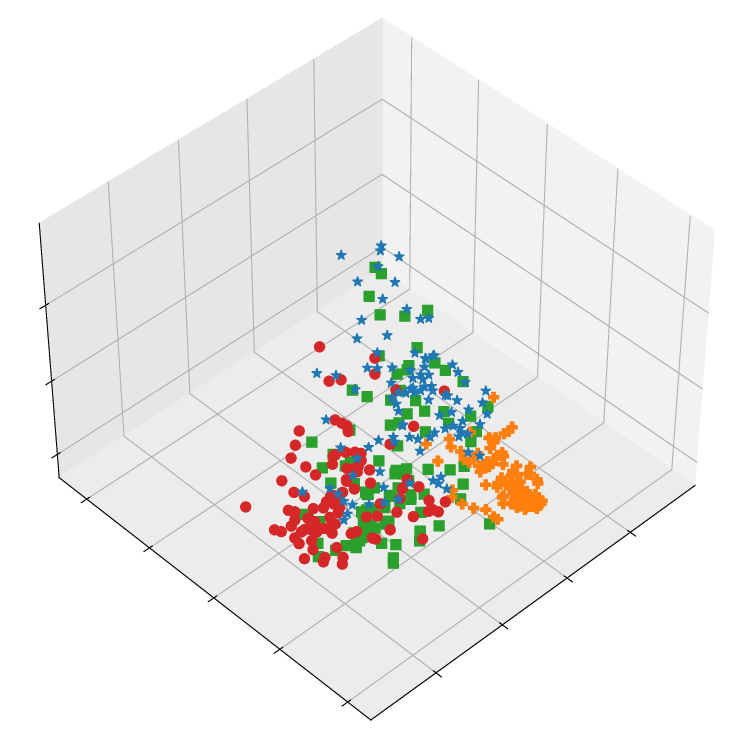}
        \caption{MobileNetV3Small}
    \end{subfigure}
    \begin{subfigure}[]{.3\linewidth}
        \centering
        \includegraphics[width=\linewidth]{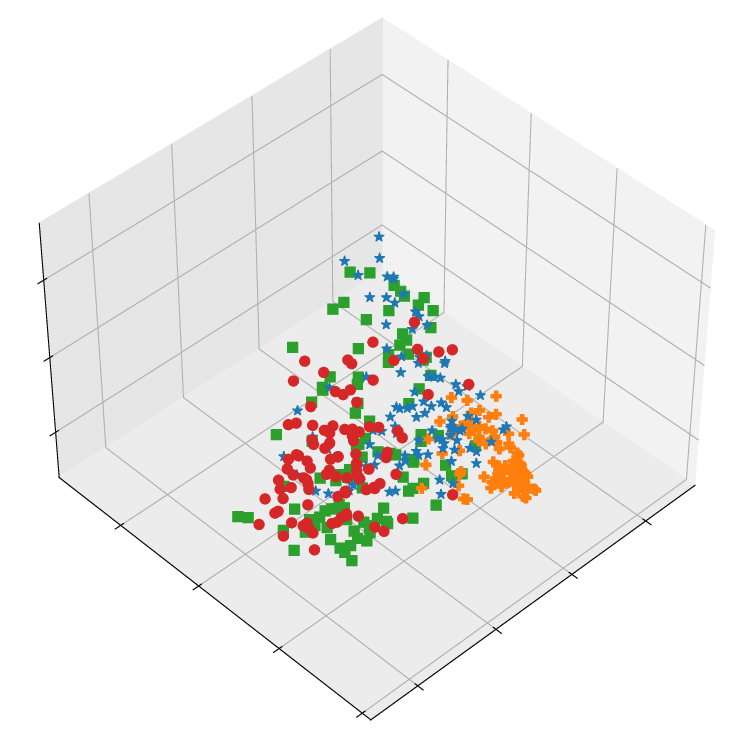}
        \caption{MobileNetV3Large}
    \end{subfigure}
    \caption{3-D representation of the dataset based on the deep embeddings learned by each model.}
    \label{fig:embeddings}
\end{figure}

\subsubsection{Deep embeddings analysis}


The obtained results are also supported by the analysis of the deep features (\textit{i.e.}, embeddings) extracted by the different CNNs.
Figure~\ref{fig:embeddings} shows how each model represents the different classes of data samples in the deep latent space, by using Principal Component Analysis (PCA) as data dimensionality algorithm to project the embeddings onto a 3-dimensional plane.

As we can note, for both InceptionResNetV2 and NASNetMobile, it is particularly difficult to distinguish the $4$ data clusters: while in the data space modeled by the former CNN, the data points are mainly concentrated in a single blob, in the latter they are distributed on a V-shaped hyperplane, where data of different classes are overlapped to each other.
By contrast, the data space modeled by VGG16 makes it easier to distinguish the different classes, even though data points belonging to \texttt{Healthy} are still considerably mixed with both \texttt{Acne} and \texttt{Mpox} samples.
The best deep representations are given by the two MobileNetV3 variants, where the considered classes are well-separated.
In addition, it is worth noting a lower data dispersion in the MobileNetV3Small embeddings space, thus facilitating the separation of the $4$ clusters and, consequently, better classification performances.

\section{Analysis of Grad-CAM indications}
\label{sec:gradcam}

\begin{figure}[t]
    \centering
    \includegraphics[width=.8\linewidth]{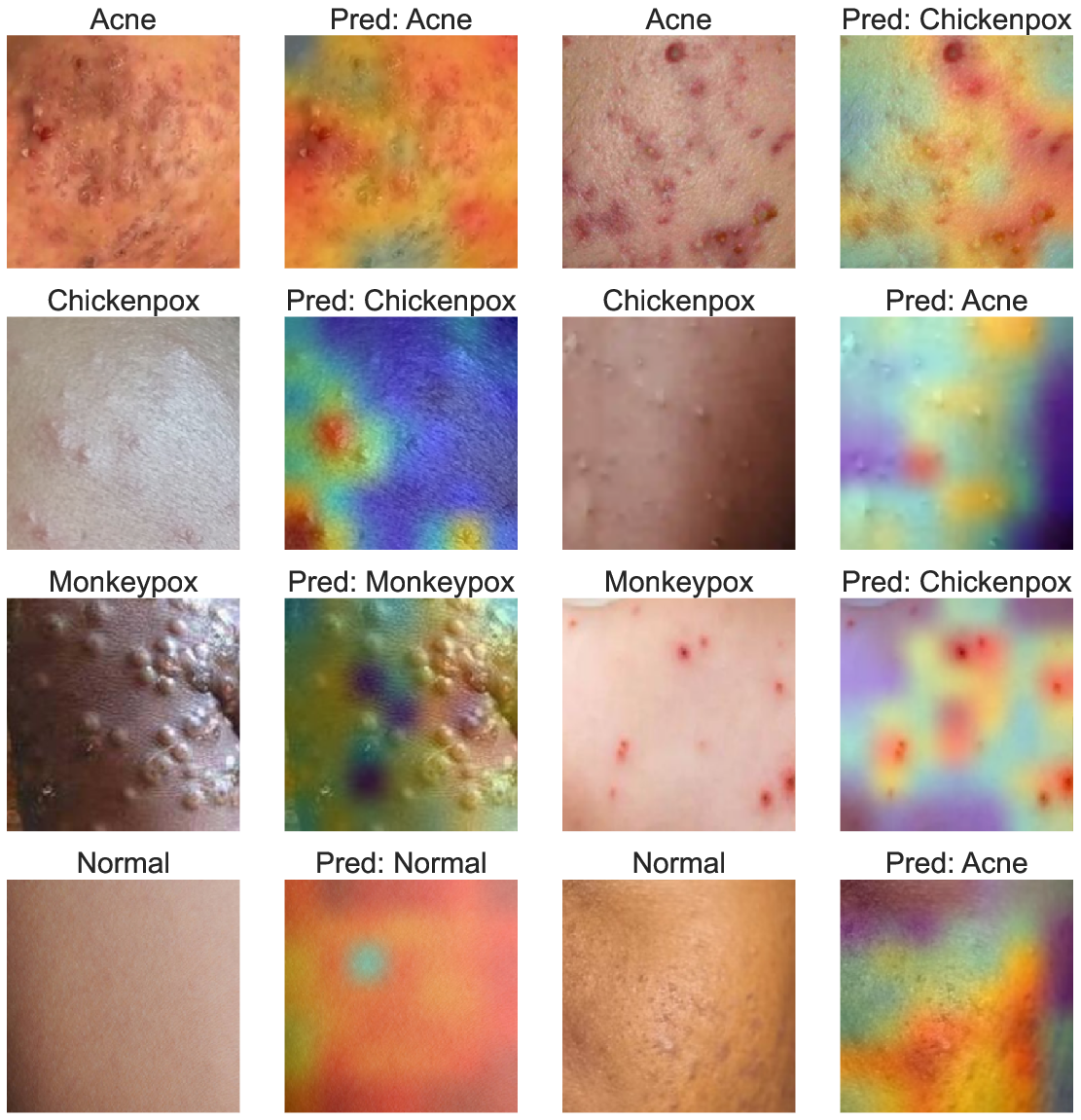}
    \caption{Examples of Grad-CAM results for each class with MobileNetV3Large, first and third columns show the input image, (Correctly and wrongly predicted respectively). Second and fourth columns show Grad-CAM explanations (for correctly and misclassified examples)}
    \label{fig:xai_examples}
\end{figure}

Grad-CAM allows us to identify the features relevant for the model for a certain prediction. For this reason, it can be used as a supporting tool to investigate the reasoning behind the model's decisions.

As one of the best-performing models in both classification tasks, we decided to apply Grad-CAM to the predictions provided by MobileNetV3Large.
Specifically, in order to understand what features of the input images are considered relevant by the model, in Fig.~\ref{fig:xai_examples} we reported $8$ different examples of explanations, four correctly predicted, along with their class activation maps (first and second columns), and four misclassified samples, with their corresponding maps (third and fourth columns).
The ground-truth label and the predicted one are indicated at the top of each image, while the heatmaps have been generated by superimposing the class activation map to the original image.
While bluish areas identify less relevant features for the given class, warmer colors (e.g., orange and red) represent the most relevant ones that have led the models to provide the specified prediction.

For example, the first row represents a case of \texttt{Acne}. When the model correctly classifies the image, the relevant features are distributed across all scars and pustules, which are typical of a strong presence of acne.
However, when the model misclassifies the image, the main focus of the network is on the pimples, neglecting the skin scars, causing the model to classify the image as \texttt{Chickenpox}.

Regarding the \texttt{Chickenpox} sample, when the model provides a correct prediction, its focus is only on the largest pimples, whereas when the model makes an incorrect prediction, its attention is distributed to minor skin defects in addition to the pimples, classifying the image as \texttt{Acne}.

For \texttt{Mpox}, the model is capable of correctly identifying the pathology when vesicles and crusts are formed, but it clearly fails in the early stages of the pathology, when pimples have not yet fully developed, providing a wrong prediction (i.e., \texttt{Chickenpox} in this case).

Finally, when the model correctly classifies a \texttt{Healthy} image, as we can expect, the importance of the feature is evenly distributed throughout the image without focusing on specific elements. On the contrary, when the model misclassifies a healthy sample, it is because it gives great relevance to hair and skin damage, classifying the image as \texttt{Acne}.

The model's visual attention analysis shows that MobileNetV3Large effectively identifies reasonable features for each class. The model's misclassifications are justifiable due to the similarity of the different classes, and, despite these errors, the model's overall ability to identify relevant features highlights its potential in our specific use-case scenario, providing more reliability on the model's predictions.

\section{Impact of mobile optimization}
\label{sec:optimization}

\begin{table}[t]
\caption{Model sizes and classification performance with mobile optimization.}
\label{tab:optimization_results}
\centering
\resizebox{\textwidth}{!}{%
    \begin{tabular}{clcrrrrr}
    \toprule
    Task & Base model & Quant. & Size (MB) & Accuracy & Sensitivity & Specificity & F-1 Score         \\
    \midrule
    \multirow{10}{*}{\rotatebox[origin=c]{90}{binary}} &
        \multirow{2}{*}{VGG16}
        & \xmark          & 268.44    & .894 $(\pm.049)$ & .841 $(\pm.075)$ & .740 $(\pm .143)$ & .851 $(\pm .070)$       \\
        &
        & \cmark  & 67.22     & .894 $(\pm.053)$ & .841 $(\pm.077)$ & .740 $(\pm .143)$ & .851 $(\pm.072)$        \\
        \cmidrule{2-8}
        
        & \multirow{2}{*}{InceptionResNetV2}
        & \xmark          & 350.53    & .735 $(\pm.056)$ & .562 $(\pm.110)$ & .220 $(\pm .266)$ & .533 $(\pm .130)$     \\
        &
        & \cmark  & 89.42     & .702 $(\pm.113)$ & .585 $(\pm.105)$ & .350 $(\pm .310)$ & .548 $(\pm.127)$    \\
        \cmidrule{2-8}
        
        & \multirow{2}{*}{NASNetMobile}
        & \xmark          & 336.37    & .830 $(\pm.036)$ & .765 $(\pm.032)$ & .640 $(\pm .070)$ & .769 $(\pm .036)$     \\
        &
        & \cmark  & 85.03     & .738 $(\pm.060)$ & .750 $(\pm.065)$ & .780 $(\pm .123)$ & .702 $(\pm.058)$    \\
        \cmidrule{2-8}
        
        & \multirow{2}{*}{MobileNetV3Small}
        & \xmark          & 211.05    & .932 $(\pm.043)$ & .883 $(\pm.070)$ & .790 $(\pm .129)$ & .902 $(\pm.064)$   \\
        &
        & \cmark  & 53.01     & .915 $(\pm.046)$ & .851 $(\pm.067)$ & .730 $(\pm .116)$ & .875 $(\pm.068)$  \\
        \cmidrule{2-8}
        
        & \multirow{2}{*}{MobileNetV3Large}
        & \xmark          & 382.93    & .928 $(\pm.042)$ & .884 $(\pm.090)$ & .800 $(\pm .200)$ & .891 $(\pm.074)$   \\
        &
        & \cmark  & 96.17     & .923 $(\pm.051)$ & .875 $(\pm.106)$ & .780 $(\pm .225)$ & .884 $(\pm.089)$  \\
    \midrule
    
    \multirow{10}{*}{\rotatebox[origin=c]{90}{multiclass}} &
        \multirow{2}{*}{VGG16}
        & \xmark & 318.21 & .779 $(\pm.053)$ & .779 $(\pm.054)$ & .927 $(\pm.018)$ & .777 $(\pm .057)$     \\
        &
        & \cmark & 79.63  & .782 $(\pm.044)$ & .782 $(\pm.044)$ & .927 $(\pm.015)$ & .779 $(\pm.050)$    \\
        \cmidrule{2-8}
        
        & \multirow{2}{*}{InceptionResNetV2}
        & \xmark & 485.12 & .398 $(\pm.088)$ & .398 $(\pm.088)$ & .799 $(\pm.027)$ & .388 $(\pm .084)$     \\
        &
        & \cmark & 122.48 & .308 $(\pm.064)$ & .306 $(\pm.065)$ & .769 $(\pm.022)$ & .243 $(\pm.075)$     \\
        \cmidrule{2-8}
        
        & \multirow{2}{*}{NASNetMobile}
        & \xmark & 259.23 & .470 $(\pm.074)$ & .470 $(\pm.074)$ & .822 $(\pm.026)$ & .467 $(\pm.076)$                     \\
        &
        & \cmark & 65.51  & .471 $(\pm.095)$ & .471 $(\pm.095)$ & .823 $(\pm.034)$ & .449 $(\pm.102)$     \\
        \cmidrule{2-8}
        
        & \multirow{2}{*}{MobileNetV3Small}
        & \xmark & 225.77 & .847 $(\pm.061)$ & .847 $(\pm.055)$ & .949 $(\pm.014)$ & .843 $(\pm.065)$     \\
        &
        & \cmark & 55.62  & .833 $(\pm.066)$ & .833 $(\pm.066)$ & .944 $(\pm.020)$ & .831 $(\pm.066)$    \\
        \cmidrule{2-8}
        
        & \multirow{2}{*}{MobileNetV3Large}
        & \xmark & 278.73 & .881 $(\pm.055)$ & .881 $(\pm.055)$ & .962 $(\pm.018)$ & .879 $(\pm.058)$   \\
        &
        & \cmark & 69.98  & .880 $(\pm.046)$ & .879 $(\pm.046)$ & .961 $(\pm.014)$ & .875 $(\pm.052)$  \\

\bottomrule
\end{tabular}
}
\end{table}


Table~\ref{tab:optimization_results} shows the great advantage of using quantization to reduce the memory footprint of the models without requiring their retraining.
As we can note, the original size of the DL models trained for mpox detection considerably varies for the different base architectures, ranging between 200 MB and almost 500 MB, which can limit their implementation on several personal mobile devices.
On the other hand, by using quantization to simply lower the operations' precision from $32$-bit floats to $16$-bit floats, all the models' sizes are reduced by approximately $4$ times.
For example, the size of VGG16 tuned for binary classification dropped from 268.44 MB to just 67.22 MB, while the size of InceptionResNetV2 for multiple classes (i.e., the most demanding model in terms of memory) has been reduced by $74.75\%$, limiting its memory footprint from 485.12 MB to 122.48 MB.

Moreover, it should be noted that quantization does not significantly affect the classification performance of most considered architectures, with an average drop of $1\%$, at most, in terms of accuracy.
Only InceptionResNetV2 and NASNetMobile suffer a higher penalization: while the former lost approximately 3\% and 9\% of accuracy in the binary and multiclass settings, respectively, the accuracy level of the latter decreased by 10\%, but only in the binary task, showing almost the same performance as the nonquantized version in the multiclass experiments.

\begin{table}[t]
\caption{Average inference times (in seconds) on different mobile devices, by using both CPU (4 threads) and GPU for the computation.}
\label{tab:mobile_exec_times}
\centering
\resizebox{\textwidth}{!}{%
    \begin{tabular}{clc|rr|rr}
    \toprule
    & & & \multicolumn{2}{c|}{Google Pixel 6a} & \multicolumn{2}{c}{Xiaomi Mi 9T} \\
    Task & Base model & Quant. & CPU & GPU & CPU & GPU  \\
    \midrule
    \multirow{10}{*}{\rotatebox[origin=c]{90}{binary}} &
        \multirow{2}{*}{VGG16}
        & \xmark  & .429 $(\pm.051)$ & .031 $(\pm.002)$ & .606 $(\pm.013)$ & .245 $(\pm.011)$ \\
        &
        & \cmark  & .104 $(\pm.013)$ & .031 $(\pm.002)$ & .430 $(\pm.021)$ & .245 $(\pm.011)$ \\
        \cmidrule{2-7}

        & \multirow{2}{*}{InceptionResNetV2}
        & \xmark  & .134 $(\pm.012)$ & .057 $(\pm.007)$ & .515 $(\pm.050)$ & .188 $(\pm.016)$ \\
        &
        & \cmark  & .064 $(\pm.005)$ & .057 $(\pm.007)$ & .441 $(\pm.039)$ & .188 $(\pm.016)$ \\
        \cmidrule{2-7}

        & \multirow{2}{*}{NASNetMobile}
        & \xmark  & .041 $(\pm.014)$ & .023 $(\pm.003)$ & .206 $(\pm.051)$ & .062 $(\pm.029)$ \\
        &
        & \cmark  & .033 $(\pm.005)$ & .023 $(\pm.003)$ & .421 $(\pm.037)$ & .060 $(\pm.027)$ \\
        \cmidrule{2-7}

        & \multirow{2}{*}{MobileNetV3Small}
        & \xmark  & .018 $(\pm.004)$ & \textbf{.011} $\mathbf{(\pm.002)}$ & .056 $(\pm.014)$ & .033 $(\pm.010)$ \\
        &
        & \cmark  & \textbf{.011} $\mathbf{(\pm.002)}$ & \textbf{.011} $\mathbf{(\pm.002)}$ & .104 $(\pm.023)$ & \textbf{.032} $\mathbf{(\pm.010)}$ \\
        \cmidrule{2-7}

        & \multirow{2}{*}{MobileNetV3Large}
        & \xmark  & .018 $(\pm.010)$ & .013 $(\pm.004)$ & .067 $(\pm.028)$ & \textbf{.032} $\mathbf{(\pm.019)}$ \\
        &
        & \cmark  & .014 $(\pm.004)$ & .013 $(\pm.003)$ & .140 $(\pm.040)$ & \textbf{.032} $\mathbf{(\pm.020)}$ \\
    \midrule
    
    \multirow{10}{*}{\rotatebox[origin=c]{90}{multiclass}} &
        \multirow{2}{*}{VGG16}
        & \xmark  & .423 $(\pm.067)$ & .031 $(\pm.002)$ & .612 $(\pm.012)$ & .249 $(\pm.012)$ \\
        &
        & \cmark  & .117 $(\pm.084)$ & .031 $(\pm.002)$ & .196 $(\pm.007)$ & .249 $(\pm.012)$ \\
        \cmidrule{2-7}

        & \multirow{2}{*}{InceptionResNetV2}
        & \xmark  & .139 $(\pm.008)$ & .059 $(\pm.008)$ & .243 $(\pm.007)$ & .192 $(\pm.020)$ \\
        &
        & \cmark  & .065 $(\pm.004)$ & .059 $(\pm.008)$ & .141 $(\pm.016)$ & .192 $(\pm.020)$ \\
        \cmidrule{2-7}

        & \multirow{2}{*}{NASNetMobile}
        & \xmark  & .036 $(\pm.009)$ & .021 $(\pm.002)$ & .084 $(\pm.022)$ & .051 $(\pm.021)$ \\
        &
        & \cmark  & .031 $(\pm.003)$ & .021 $(\pm.003)$ & .127 $(\pm.015)$ & .052 $(\pm.021)$ \\
        \cmidrule{2-7}

        & \multirow{2}{*}{MobileNetV3Small}
        & \xmark  & .011 $(\pm.007)$ & .009 $(\pm.003)$ & .028 $(\pm.016)$ & \textbf{.024} $\mathbf{(\pm.015)}$ \\
        &
        & \cmark  & \textbf{.008} $\mathbf{(\pm.003)}$ & .009 $(\pm.003)$ & .040 $(\pm.009)$ & .040 $(\pm.009)$ \\
        \cmidrule{2-7}

        & \multirow{2}{*}{MobileNetV3Large}
        & \xmark  & .016 $(\pm.005)$ & .012 $(\pm.001)$ & .047 $(\pm.014)$ & .029 $(\pm.007)$ \\
        &
        & \cmark  & .014 $(\pm.002)$ & .012 $(\pm.001)$ & .062 $(\pm.004)$ & .029 $(\pm.007)$ \\

    \bottomrule
    \end{tabular}
}
\end{table}

Besides the memory size and classification performance, we also conduct an empirical evaluation of the models' time complexity.
Even though our application scenario does not require real-time predictions, fast computation represents a key requirement when dealing with mobile personal devices like smartphones.
Therefore, to perform this type of experiment, we rely on the benchmark tool provided by TensorFlow Lite (TFLite)~\footnote{\url{https://www.tensorflow.org/lite}}, the Google-released mobile library for deploying models on mobile devices, microcontrollers, and other edge devices.
Specifically, we first convert our CNN models to the TFLite format; then, we deploy such models on the TFLite Android benchmark app\footnote{\url{https://www.tensorflow.org/lite/performance/measurement##benchmark_tools}} that executes each model 50 times with synthetic input to collect reliable statistics related to the inference times on a real Android smartphone.
Moreover, in order to get insights on the models' performance on different hardware settings, we perform our evaluation on 2 smartphones, by using both CPU (with multithreading) and GPU for the computation: (i) a recent Google Pixel 6a released in 2022, with the latest Android 13 operating system, an Octa-Core CPU (2x2.80 GHz Cortex-X1, 2x2.25 GHz Cortex-A76, and 4x1.80 GHz Cortex-A55), and the Mali-G78 MP20 GPU; and (ii) an older Xiaomi Mi 9T, released in 2019, with Android 10, an Octa-core CPU (2x2.2 GHz Kryo 470 Gold and 6x1.8 GHz Kryo 470 Silver), and an Adreno 618 GPU.

Table~\ref{tab:mobile_exec_times} summarizes the average inference times (in seconds) of the considered models in the different hardware settings, both for the binary and multiclass classification tasks, highlighting in bold face the best results for each device and task.
It is clear that even the largest models such as VGG16 and InceptionResNetV2 can provide a prediction in less than 0.612 seconds when deployed on modern smartphones.
The benefit of using quantization can be mainly observed when the computation is based on CPU, reducing the inference time by 50\% at least in some cases (e.g., VGG16 and InceptionResNetV2 with Google Pixel 6a).
On the other hand, all models can be executed by the GPU in less than 0.059 seconds on Google Pixel 6a and 0.245 seconds on Xiaomi Mi 9T, thanks to its ability to parallelize all operations that are involved in a deep neural network~\cite{7573804}.

Finally, we can also note that the CNN that performs best in terms of classification accuracy, i.e., MobileNetV3 (both Small and Large variants), is also the one with the lowest inference time.
In fact, while the larger variant provides a prediction for binary and multiclass classification, respectively, in not more than 0.018 and 0.016 seconds on Google Pixel 6a and not more than 0.140 and 0.062 seconds with Xiaomi Mi 9T, MobileNetV3Small requires only not more than 0.018 and 0.011 seconds on the Google phone and not more than 0.104 and 0.040 seconds on the Xiaomi, thus proving the feasibility of efficiently performing the whole data processing and prediction tasks directly on mobile devices.



\section{Conclusions and future work}
\label{sec:conclusions}

The paper introduces a novel mobile health (m-health) system for the preliminary screening of mpox infections through pictures of skin rashes and eruptions taken with common smartphone cameras.
The system is designed to be entirely executed on mobile devices and is characterized by the use of Transfer Learning to adapt state-of-the-art Convolutional Neural Network (CNN) models for image classification, mobile-oriented optimization of the models through quantization, and the use of Grad-CAM as eXplainable AI (XAI) technique for technical validation.

We also presented a homogenous, unpolluted dataset derived from the manual selection and preprocessing of available images data of skin lesions collected from crowsourcing projects. 
We called the resulting dataset Mpox Close Skin Images (MCSI), and we used it to evaluate the classification performance of the proposed system, using both binary (mpox vs. all) and multiclass classification tasks with a 10-fold stratified cross-validation approach.
The results showed that MobileNetV3Small achieved the best performance in binary classification (0.930 of Accuracy), while MobileNetV3Large was the best model to distinguish the different classes (0.882 of Accuracy).
Mobile optimization through quantization allowed us to reduce the models' sizes by $4\times$ without significantly impacting their performance.
The models have also been evaluated for their complexity in terms of execution time on commercial smartphones, and they all obtained performances under 1 second to provide the prediction, with quantization further reducing the inference time on CPUs.

Despite achieving promising results, the primary limitation of our study arises from the paucity of public training data.
The performance of a CNN model is highly dependent on the availability of high-quality training samples, which is a critical factor.
Therefore, in order to enhance the performance and reliability of the proposed m-health system, we intend to collaborate with healthcare experts, such as dermatologists or virologists, to gather a more diverse and comprehensive set of skin lesion images.
This would entail creating a vast dataset that covers various populations, ethnicities, and age groups.
To accomplish this objective, we aim to develop a prototype mobile application that can not only be used for data collection on the field, but also to deploy the mpox detection system in real-world settings.
This would enable us to evaluate its feasibility, usability, and acceptability with the assistance of medical professionals.

Future research includes investigating the Federated Learning (FL) technique in this use-case scenario, which has the potential to improve the m-health system in various ways.
First, FL facilitates the collaborative training of the detection model by mobile devices without the need to share users' data with a central server or other devices.
Each device gathers data locally, trains a model using it, and then shares only the model updates with a central server, which aggregates and distributes them back to all the devices.
This approach can address privacy concerns since sensitive health data remains under the user's control. Secondly, training the model with data from multiple devices can improve the accuracy of the model by incorporating more diverse data.
This is particularly crucial for the detection of skin lesion, where the types of lesion and the color of the skin can vary significantly between various populations.
Finally, FL can enable real-time updates of the detection model as new data becomes available, thus aiding the system to adapt to data changes and further enhance the model's accuracy over time.

\section*{Acknowledgment}

This work was produced with the co-funding European Union - Next Generation EU, in the context of The National Recovery and Resilience Plan. The funding derives partially from Investment 1.5 Ecosystems of Innovation, Project Tuscany Health Ecosystem (THE), CUP: B83C22003920001 in which the authors M. G. Campana and F. Delmastro are involved, from Project MUSA – Multilayered Urban Sustainability Action in the Investment 1.5 Ecosystems of Innovation in which the author S. Mascetti is involved, and from the Research and Innovation Program PE00000014,  ``SEcurity and RIghts in the CyberSpace (SERICS)'', CUP: XXX, in which the author E. Pagani is involved. 

.

\bibliography{paper}

\end{document}